\documentclass[preprint]{aastex63}
\usepackage{graphicx}
\graphicspath{{figures/}}
\usepackage{amsmath}
\usepackage{lineno}
\usepackage{makecell}
\usepackage{amssymb}
\usepackage{multirow}
\usepackage{multirow}
\usepackage{color}

\shorttitle{Psyche Millimeter Emission}
\shortauthors{de Kleer et al.}

\begin{document}

\title{The Surface of (16) Psyche from Thermal Emission and Polarization Mapping}
\correspondingauthor{Katherine de Kleer}
\email{dekleer@caltech.edu}

\author{Katherine de Kleer}
\affil{California Institute of Technology \\
1200 E California Blvd M/C 150-21 \\
Pasadena, CA 91125, USA}
\author{Saverio Cambioni}
\affil{California Institute of Technology \\
1200 E California Blvd M/C 150-21 \\
Pasadena, CA 91125, USA}
\author{Michael Shepard}
\affil{Bloomsburg University \\
Bloomsburg, PA 17815, USA}

\begin{abstract}
The asteroid (16) Psyche is the largest of the M-type asteroids, which have been hypothesized to be the cores of disrupted planetesimals and the parent bodies of the iron meteorites. While recent evidence has collected against a pure metal composition for Psyche, its spectrum and radar properties remain anomalous. We observed (16) Psyche in thermal emission with the Atacama Large (sub-)Millimeter Array (ALMA) at a resolution of 30 km over 2/3 of its rotation. The diurnal temperature variations are at the $\sim$10 K level over most of the surface and are best fit by a smooth surface with a thermal inertia of 280$\pm$100 J m$^{-2}$ K$^{-1}$ s$^{-1/2}$. We measure a millimeter emissivity of 0.61$\pm$0.02, which we interpret via a model that treats the surface as a porous mixture of silicates and metals, where the latter may take the form of iron sulfides/oxides or alternatively as conducting metallic inclusions. The emissivity indicates a metal content of no less than 20\% and potentially much higher, but the polarized emission that should be present for a surface with $\geq$20\% metal content is almost completely absent. This requires a highly scattering surface, which may be due to the presence of reflective metallic inclusions. If such is the case, a consequence is that metal-rich asteroids may produce less polarized emission than metal-poor asteroids, exactly the opposite prediction from standard theory, arising from the dominance of scattering over the bulk material properties. 
\end{abstract}


\section{Introduction} \label{sec:intro}

Our Solar System's population of asteroids represents the remnant fragments of the planetesimals, and encodes the history of the thermal and collisional evolution of the inner Solar System \citep[e.g.][]{wetherill1988}. Objects the size of the large asteroids that formed early should have trapped enough heat during formation to differentiate \citep{dauphas2011}, and some of the crust, mantle, and core fragments of these objects must remain today in the asteroid belt. The M-type asteroids (\citealt{tholen1984}; now a subset of the X complex, \citealt{bus2002,demeo2009}) are hypothesized to be the remnant core fragments of these differentiated objects, and the parent bodies of the iron meteorites \citep{chapman1973,johnson1973,gaffey1978,bell1989}. The collection of iron meteorites is thought to originate from at least $\sim$75 unique parent bodies \citep{keil1994,meibom1999}, implying that numerous metallic core fragments should be present in the asteroid belt \citep[e.g.,][]{2015asteScott}. \par
The M-type asteroids show red sloped, nearly featureless optical-near-infrared spectra, which are characteristic of the iron meteorites \citep{cloutis1990}. These asteroids are also characterized by high radar albedos, which imply higher surface bulk densities than typical asteroids. The object (16) Psyche (hereafter given without its numerical designation) is the largest asteroid in this class; its radar albedo of 0.34$\pm$0.08 \citep{shepard2021} implies a surface bulk density around 3.5 g/cm$^3$, which is consistent with a metallic surface of 50-60\% porosity, or alternatively with a low-porosity surface of enstatite chondrite composition \citep{ostro1985,shepard2017}. Psyche's bulk density of 3.78$\pm$0.34 g/cm$^3$ \citep{elkins-tanton2020} is consistent with its surface density, implying minimal increase in density with depth.\par
The presence of shallow near-infrared spectroscopic features demonstrates that whether or not Psyche's bulk composition is predominantly metal, its surface composition is a mix of metals, silicates, and hydrated minerals. The presence of the 0.9-$\mu$m pyroxene band over Psyche's entire rotation indicates the presence of orthopyroxene across its surface, although the feature depth is only 1-3\% and varying spatially  \citep{hardersen2005,sanchez2017}. The presence of this band is inconsistent with the silicates in enstatite chondrites, whose iron-free compositions do not produce this spectral feature \citep{hardersen2005}. Detection of the 3 $\mu$m absorption feature in Psyche's spectrum also indicates that its surface hosts OH- or H$_2$O-bearing phases \citep{takir2017}; the sharp shape of the feature indicates that these take the form of phyllosilicates \citep{takir2012}. The presence of such species suggests either that Psyche is not a metallic core fragment, or that the hydrated species are exogenic \citep{avdellidou2018}. Recent ultraviolet observations show that Psyche's UV spectrum is most consistent with that of iron, though even modest amounts of iron ($\sim$10\%) can produce the observed spectral shape \citep{becker2020}. This is also the case in the optical-near-infrared, where both enstatite chondrites and particulate iron meteorite material match Psyche's spectrum well \citep{johnson1973,chapman1973} because in both cases the spectrum is dominated by its metal component \citep{gaffey1976}. Collectively studies to date have been unable to unambiguously discriminate between the porous metallic core fragment and the low-porosity enstatite chondrite hypotheses, and the composition and origin of Psyche, and the M-type asteroid class as a whole, remains inconclusive.\par
Thermal emission observations provide constraints on thermophysical properties such as thermal conductivity, and hence complement the radar and reflectance spectroscopy studies. Currently, two discrepant findings exist for Psyche's thermal inertia, both based on mid-IR observations. \cite{matter2013} found evidence for a surface with little or no roughness and a thermal inertia $\Gamma$ in roughly the 100-150 J m$^{-2}$ K$^{-1}$ s$^{-1/2}$ range ($\Gamma$ will be given unitless henceforth but always refer to these SI units), much lower than solid rock but higher than most large asteroids, while \cite{landsman2018} found a thermal inertia in the 5-25 range and inferred the presence of a fine-grained silicate regolith. Observations at millimeter wavelengths and longer are also sensitive to thermal emission, but detect emission from the shallow subsurface and provide constraints on the electrical properties of the material in addition to the thermal properties. At these wavelengths, the intensity and polarization of emission are influenced by the complex dielectric constant of the material, in addition to the roughness and scattering properties of the surface. The complex dielectric constant is particularly sensitive to the metal content of a surface: metal-rich objects should exhibit low emissivities and high polarization fractions, and measurements of the emission properties of a surface at millimeter wavelengths can therefore provide constraints on a surface's metal content.\par
Spatially resolved thermal emission measurements can further differentiate between candidate scenarios for surface characteristics. Direct observations of how the brightness temperature varies across the surface can disentangle the roles of thermal properties, which set the actual surface temperature, and electrical properties, which set the degree of emission relative to a blackbody. In addition, the polarization of the emission, in particular when its variation with emission angle can be resolved, provides an independent constraint on the dielectric constant and on the degree of scattering.\par
Mapping the thermal emission from asteroids $\sim$200 km in size has not been possible historically due to the very high angular resolution required ($\sim$0.03''), which is not achievable at mid-infrared thermal wavelengths even with a 10 meter telescope. In addition, the polarization of thermal emission from asteroids has only been measured in the infrared, and only in disk-integrated observations \cite[e.g.][]{johnson1983}, in which the polarization vectors largely cancel. The Atacama Large (sub)Millimeter Array (ALMA) is now enabling these measurements for the first time, providing a new technique for studying the surface properties of asteroids. Thermal mapping of asteroids with ALMA has only been performed for (3) Juno to date, and the observations did not measure the polarization of the emission \citep{junoalma}. \par
Using ALMA, we mapped the intensity and polarization of the 1.3-mm thermal emission from Psyche at a spatial resolution of $\sim$30 km (0.02'') over 2/3 of its rotation (Section \ref{sec:data}). We interpret the results in terms of the thermal, electrical, and scattering properties of the surface; the theory is presented in Section \ref{sec:models}, and results in Section \ref{sec:results}. In Section \ref{sec:disc} these results are placed in context of previous data at multiple wavelengths. The conclusions are summarized and interpreted in terms of candidate scenarios for Psyche's surface composition in Section \ref{sec:conc}. This study constitutes the first spatially resolved thermal observations of an M-type asteroid at this high resolution, and the first millimeter polarization map of an asteroid's thermal emission. As such, it also represents a demonstration of a capability that can be applied to the asteroid belt more broadly.\par
\section{Data reduction, calibration, and imaging} \label{sec:data}
Observations were made of Psyche with ALMA, located in the Atacama Desert in Chile, on UT 2019 June 19 between 06:33 and 09:11 UT. ALMA was in an extended configuration at the time of observation, with a maximum baseline of 16.2 km resulting in an angular resolution of $\sim$0.02''. This corresponds to $\sim$30 km on Psyche, which was at a distance of 2.78 AU from the Sun and 2.04 AU from Earth at the time of observation and subtended $\sim$0.15''. Data were collected in continuum mode with four spectral windows each containing 64 channels over 2 GHz of bandwidth, centered at 224, 226, 240, and 242 GHz (around 1.3 mm, ALMA's Band 6). Psyche was observed alternately with calibrators; the on-source time consisted of 88 scans each of duration 55 seconds or less, for a total on-source time of 73 minutes. Quasar J2131-1207 was used for phase calibration, J2134-0153 for polarization calibration, and J2258-2758 for bandpass calibration and flux density scale calibration. As a check on the flux density scale calibration, the flux density of J2134-0153 as derived from the flux density scale calibrator was compared against direct measurements from the ALMA calibrator source catalogue, and was found to correspond well. For this reason, as well as the stability of J2258-2758 during the months surrounding the observations, we adopt a flux density scale calibration uncertainty of 3\%. A larger uncertainty would translate directly into a larger uncertainty on our measured millimeter emissivity, but would not qualitatively affect the results.  \par
The raw data were reduced and calibrated via the ALMA pipeline; the data were delivered to us in the form of a calibrated Measurement Set (MS). The MS contains, for each target, the amplitude and phase of the cross-correlated signal between each antenna pair. These quantities, referred to as ``visibilities,'' are the fundamental measured quantities of any radio interferometer, and are a sampling of the Fourier transform of the sky brightness temperature distribution at discrete spatial frequencies. Processing of the MS was performed using the Common Astronomy Software Applications (CASA) package \citep{mcmullin2007}. \par
We performed an iterative imaging and self-calibration procedure \citep{cornwell1999} on the visibilities to improve the phase calibration and produce an image for comparison with models. Imaging was performed with the CASA task \textbf{tclean} using a multi-scale multi-frequency algorithm for deconvolution \citep{rau2011}. Individual scans have a signal-to-noise (SNR) in the $\sim$15-25 range, and self-calibration on individual scans was unsuccessful. Instead, self-calibration was performed on sets of 3-5 scans together, where the total duration of the 3-5 scans never exceeded 6 minutes of clock time to minimize smearing due to Psyche's rotation (the asteroid rotates through about half a beam at target center in 6 minutes). Each set of scans was imaged first using a shallow clean, and the resultant image was used as the input for self-calibration. Phase-only self calibration was performed with a solution interval equal to the entire observing duration of the set of scans being imaged. Further iterations of self-calibration with shorter solution intervals did not improve the SNR and were therefore not employed. The self-calibration procedure was additionally tested using a thermal model image of Psyche as the input model, but this approach was ultimately not used because it did not substantially improve SNR and it biased the resulting image towards flux density distribution of the input model. Given that Psyche's thermal properties are imperfectly known, we prefer the approach that does not bias the results towards previous assumptions. The self-calibrated sets of scans were then imaged using a deeper clean to produce the final images. The self-calibration procedure ultimately improved the SNR by $\sim$20\%, resulting in a SNR of $\sim$45-50 for each image and a typical image noise level of 2.5 K. The final images have a pixel scale of 3 milliarcseconds (mas) and an angular resolution of 19-22 mas. \par
\begin{figure}[ht]
\centering
\includegraphics[width=16cm]{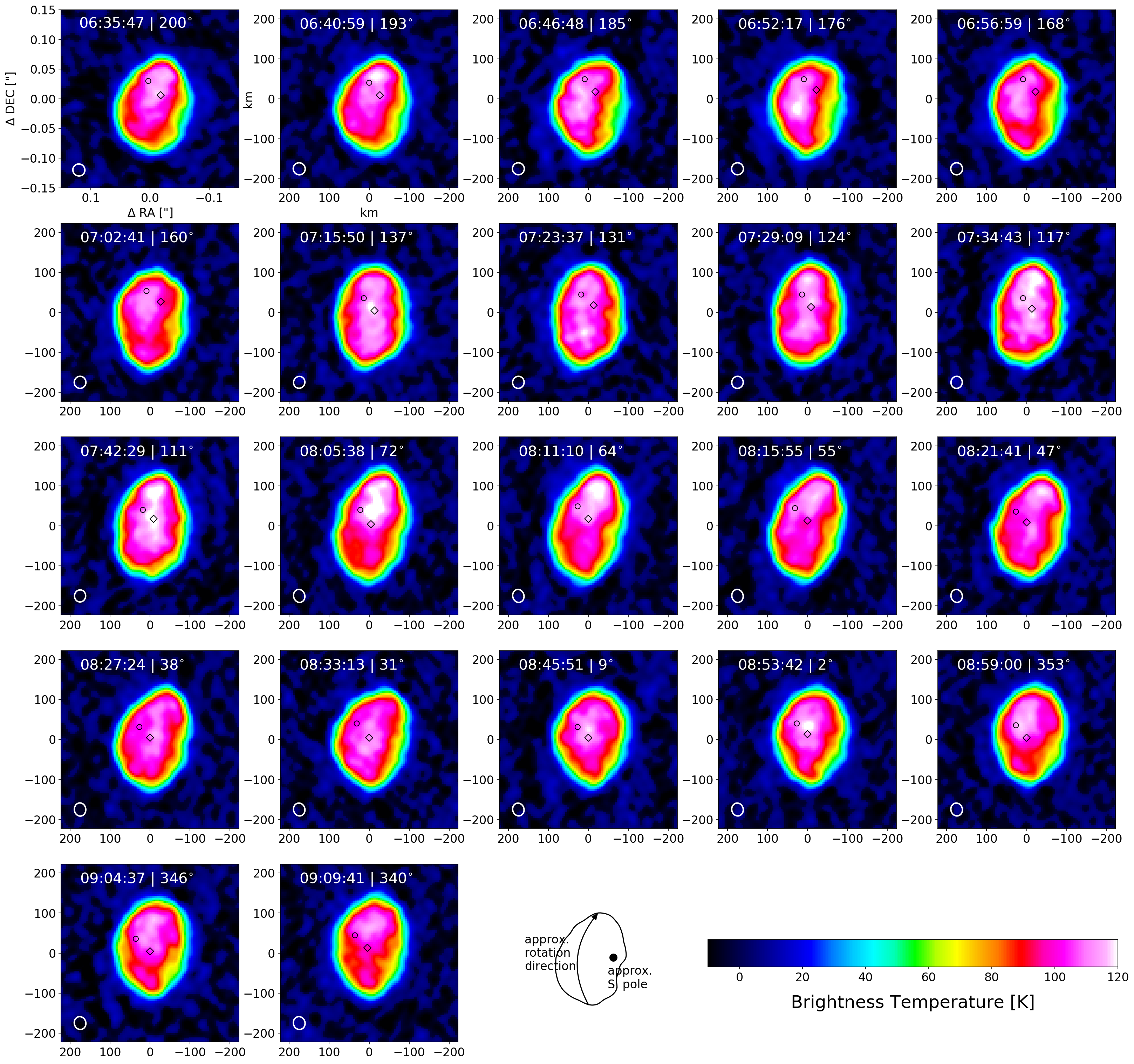}
\caption{ALMA images of Psyche over $\sim$2/3 of its rotation period. The spatial and intensity scales are the same across all images; the spatial scale is given in units of arcseconds in the first panel and km in the following panels. The ellipse in the lower left corner represents the resolution element. The UT time and sub-observer longitude are given for each image, and the sub-solar and sub-observer points (based on the \cite{shepard2021} shape model) are shown as a small circle and diamond respectively. The sub-solar and sub-observer latitudes are 3$^{\circ}$ and -14$^{\circ}$ respectively, and the solar phase angle is 17$^{\circ}$. An animation version of the images in this figure is provided with the paper.\label{fig:ims}}
\end{figure}
\begin{figure}[ht]
\centering
\includegraphics[width=8cm]{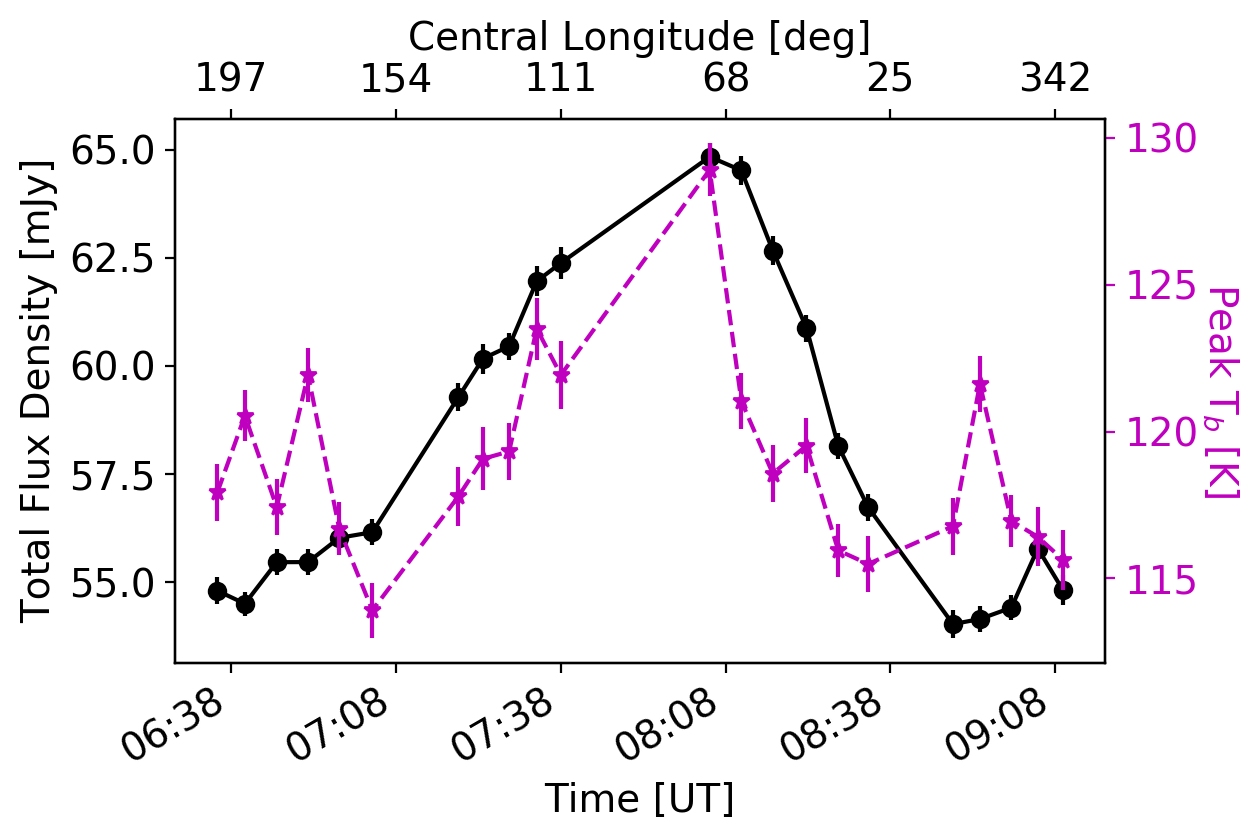}
\caption{Total flux density (solid lines, circles) and peak brightness temperature (dashed lines, stars) for each ALMA observation, as a function of time and sub-observer longitude. The uncertainties are appropriate for comparison between observations, but do not include the flux density scale calibration uncertainty of 3\% that applies to the absolute calibration of the dataset as a whole.\label{fig:diskint}}
\end{figure}

Each set of jointly-imaged scans is labeled as a single observation in Table \ref{tbl:obs}, which provides further details on the observations. Table \ref{tbl:obs} also presents the integrated flux density of Psyche in each observation, which is calculated by summing the flux density in all pixels within a radius of 0.15'' of the target center that have a flux density greater than 3$\sigma$, with $\sigma$ defined as the root mean square (rms) of the image background. The uncertainty on this value is determined by propagating the image rms through the number of resolution elements (beams) across the target; the 3$\sigma$ uncertainty is presented in Table \ref{tbl:obs}. The disk-averaged brightness temperature $<T_b>$ is calculated by taking the median flux density of the pixels in the clean model that fall on Psyche's disk according to the \cite{shepard2021} shape model projected to the time of observation, and converting the flux density to a brightness temperature using the Planck function and the solid angle subtended by a pixel. The peak brightness temperature in each observation $T_{b,peak}$ and its uncertainty are computed from the peak flux density and image rms using the Planck function. The final set of calibrated images is shown in Figure \ref{fig:ims}, and Figure \ref{fig:diskint} plots the disk-integrated brightness and $T_{b,peak}$ for each observation as a function of time and sub-observer longitude.\par
\begin{figure}[ht]
\centering
\includegraphics[width=8cm]{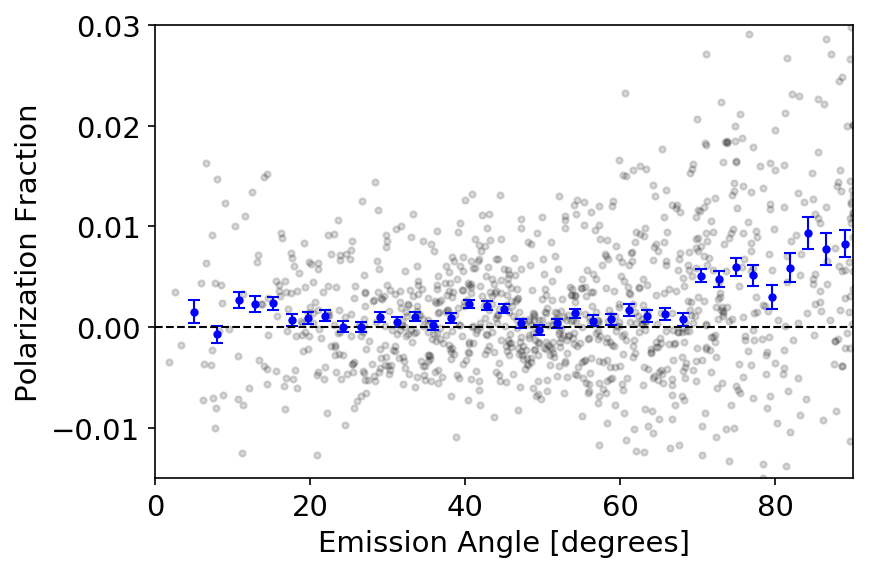}
\caption{Fractional polarization as a function of emission angle on Psyche. The gray dots are datapoints from each scan, while the blue points are binned across the entire observation.\label{fig:polfrac}}
\end{figure}

The linearly polarized emission $P$ is computed from the Stokes $Q$ and $U$ parameters as $P=\sqrt{Q^2+U^2}$. Polarization images were produced for each set of scans, and the images were averaged together to produce the final image shown in Figure \ref{fig:polimage}, which is also smoothed to lower spatial resolution. The polarized emission is only detected from the lower half of Psyche, and even that at very low SNR. The averaging of polarization images circumvents the cancellation of polarization vectors that would occur in imaging of the entire observation jointly. Figure \ref{fig:polfrac} shows the polarization fraction ($P/I$ for the Stokes $I$ parameter) as a function of emission angle, averaged across all observations, demonstrating that the polarization fraction increases towards the limb as expected. Emission angles are calculated from the \cite{shepard2021} shape model, which was derived from a large set of observations including those presented here, and hence provides an excellent match to our images. The errorbars on the binned points in Figure \ref{fig:polfrac} are derived from the noise in the polarization images, propagating based on the number of datapoints in each emission angle bin. \par
\section{Thermal and electrical modeling}\label{sec:models}
\subsection{Approximating peak surface temperature}\label{sec:stm}\label{sec:thermal}
We first calculate bounds on the expected surface temperature of Psyche by approximating the full plausible range of subsolar temperatures using the Standard Thermal Model  \citep[STM;][]{lebofsky1986} and Fast Rotating Model \citep[FRM;][]{lebofsky1989}. The STM and FRM represent the non-rotating (or zero thermal inertia) and fast-rotating (or infinite thermal inertia) end members, respectively, and can be considered the absolute upper and lower limits on the true peak surface temperature of the object. Because Psyche is resolved in our observations, we can compare the subsolar temperature predicted by these models directly to the peak temperature in the ALMA dataset, without making assumptions about the temperature distribution across the rest of the surface. In the STM, the subsolar temperature $T_{ss}$ is given by
\begin{equation}
T_{ss}=\bigg[\frac{(1-A)S}{\eta E_{bol} \sigma}\bigg]^{1/4}
\label{eqn:Tss}
\end{equation}
where $A=pq$ is the Bond albedo, which is the product of the geometric albedo $p$ and the phase integral $q$. $S$ is the incident solar flux density, which is calculated from the solar flux at 1 AU scaled to the asteroid-Sun distance at the time of observation (2.78 AU); $\eta$ is the beaming parameter; $E_{bol}$ is the bolometric emissivity; and $\sigma$ is the Stefan-Boltzmann constant. Psyche's geometric albedo was found to be 0.16 by \cite{shepard2021} and 0.12-0.13 for the best-fit thermal models of \cite{matter2013}. According to the albedo - phase integral relation given by \cite{shevchenko2019}, a phase integral of 0.38 is appropriate for an object of this albedo. A bolometric emissivity of 0.9 is appropriate for a silicate-dominated surface, whereas a lower emissivity is more appropriate for a metallic surface \citep{matter2013}. We adopt a geometric albedo of 0.12-0.16, a bolometric emissivity of 0.8-0.9, and a beaming parameter of 0.75-1.0 to calculate the range of expected $T_{ss}$ from the STM. Note that using a range of beaming parameters in the STM calculation represents a slight extension of the exact model presented in \cite{lebofsky1986}, which had a fixed beaming parameter of 0.756. For the FRM, $\eta$ is replaced with $\pi$ in Equation \ref{eqn:Tss}; the resultant $T_{ss}$ approximates the temperature at a depth where diurnal temperature variations are absent and hence can be considered a lower bound on the peak surface temperature. \par
\subsection{Thermophysical modeling of (sub)surface temperature}\label{sec:TPM}
Our observations resolve Psyche spatially and temporally over more than half its rotation, and the information content of the data is much higher than can be recovered using the simple approximations above, motivating more detailed modeling. We interpret the data using a thermophysical model that we have adapted to model spatially-resolved millimeter observations, including projecting the modeled temperatures of the asteroid facets onto the plane of the sky, and integrating emission from the subsurface as is appropriate for millimeter (and longer) wavelengths. The modeling approach is as follows.

We use a well-established thermophysical model \citep[TPM,][]{delbo2015} to build a look-up table of thermophysical solutions for the thermal inertia of the (sub)surface material $\Gamma$ varying between 25 and 1000 (specifically: 25,  75, 116,  135, 156, 181,  210, 283, 442,  594,  1000) and for surface roughness parameter $f_c$ between 0 and 0.6 (specifically, 0, 0.2, 0.4, 0.6, with 0 indicating a smooth surface). The parameter $f_c$ is the surface density of hemispherical craters carved on each triangular facet of the shape model from \citet{shepard2021}, which we use to model the asteroid topography, pole orientation and rotation period. Adding roughness to the topography allows modeling surface irregularities that are not captured by the shape model but still affect the asteroid temperature through self-heating. The thermophysical model has a surface material with bolometric emissivity of $E_{bol} =$ 0.9 and a Bond albedo of 0.053, both of which are treated as constant over the surface.\par
For every combination of surface properties, we illuminate the asteroid with sunlight and evolve its rotation for 15 Julian days to achieve stability and convergence of the diurnal temperature curve (thermalization). The light-time corrected ephemerides corresponding to a specific observation are obtained by querying the \href{https://ssd.jpl.nasa.gov/?horizons}{JPL Horizons ephemeris system}. The 1-D temperature profile in the subsurface extends down to several times the diurnal thermal skin depth $\delta_{th}$, that is, the e-folding depth of the diurnal heat wave:

\begin{equation}
    \delta_{th} = \sqrt{\frac{P}{\pi}}\frac{\Gamma}{\rho c_p}
\end{equation}
\noindent
where $\rho$ = 3.5 $g/cm^3$ is the density of the near-surface material inferred from Psyche's radar albedo \citep{shepard2021} and $c_p$ = 370 J/Kg/K is the heat capacity computed as the average values for the possible meteorite analogues for Psyche identified in \citet{elkins-tanton2020}:
mesosiderites ($c_p$ = 383 $\pm$ 6 J/Kg/K), iron IIAB ($c_p$ = 342 $\pm$ 6 J/Kg/K) and iron IIIAB ($c_p$ = 375 $\pm$ 23 J/Kg/K) at around 200 K \citep{2013Consolmagno,2013ConsolmagnoSchaefer}.

\subsection{Radiative transfer and projection onto the plane of sky}\label{sec:RT}
The thermophysical model provides the temperature profile in the subsurface as a function of time of day for each facet in the asteroid shape model. From this, we compute the model emission from each facet by integrating the thermal emission from each subsurface layer following the approach outlined in \citet{dekleer2021}. We integrate the subsurface emission for 17 values of real dielectric constant of the material $\epsilon'$ ranging between 3 to 80.

The thermal emission from each facet is then corrected for the millimeter emissivity of the surface $E = E(\theta)$, which is computed from $\epsilon'$ as will be described in Section \ref{sec:electrical}. For the smooth-surface model runs, $E(\theta)$ follows the Fresnel equations and is given by Equation \ref{eqn:E}. When the TPM is run with surface roughness, $E(\theta)$ follows the rough surface model which will be described in Section \ref{sec:defn}. The roughness emissivity model is parameterized by the adirectional rms slope angle $s$ \citep{spencer1990}. We use the relation $s = 49\sqrt{f_c}$ degrees \citep{rozitis2020asteroid} to convert between the emissivity roughness parameter $s$ and the TPM roughness parameter $f_c$, to ensure internal consistency in the model.\par
The TPM has been traditionally used to interpret disk-integrated observations. To interpret the spatially resolved ALMA observations, we identify the facets of the shape model that are visible to an observer located at ALMA at the acquisition time of each observation, and project them onto the plane-of-sky to obtain a model image in right ascension and declination, with the same pixel size as the images. We first compute the ICRS coordinates of the facets projected onto the plane of the sky; the transformation from the asteroid reference frame to the ICRS follows the formalism introduced by \cite{2010Durech}. We then interpolate the projected facets and the corresponding thermal emission values in right ascension and declination to produce a 2D array whose pixels correspond to those of the ALMA images. Finally, we convolve the model images with the ALMA beam for comparison with the data.

\subsection{Goodness-of-fit and uncertainty estimation}

We estimate the goodness-of-fit of the simulations by means of the cost function

\begin{equation}
\label{eq:chi2}
    \chi^2_{r}(\Gamma,~f_c,~\epsilon') = \frac{1}{N_{pxl} - N_{par}} \sum_{i = 1}^{N_{obs}} \sum_{j=1}^{N_{pxl}^i} \bigg(\frac{(M_j^i - D_j^i)}{\sigma^i}\bigg)^2
\end{equation}
\noindent
where $M_j^i$ and $D_j^i$ are the modeled and observed flux densities (respectively) of the $j$th pixel that is on the target, in the $i$th observation. $N_{pxl}$ is the total number of pixels across Psyche, summed across all observations, and the number of free parameters $N_{par}$ is 3 here. The noise level $\sigma^i$ associated with the $i$th observation is the image rms described in Section \ref{sec:data}. We acknowledge that, formally, the cost function should treat independent datapoints, whereas the number of pixels is much larger than the actual number of resolution elements (as is also the case in fitting thermal models to unresolved spectra, where this approach is frequently employed). 
We therefore adopt a conservative estimate of the uncertainties, based on the number of independent datapoints rather than the numbers of pixels. The uncertainties on the surface properties ($\Gamma$, $f_c$, $\epsilon'$) are estimated by means of the $\chi^2_r$ statistics as done in previous studies \citep[e.g.,][]{hanuvs2015thermophysical,cambioni2019constraining}, that is, by considering that every combination of ($\Gamma$, $f_c$, $\epsilon'$) which satisfies $\chi^2_r < \chi^2_{r,min}\big(1 + \sqrt{2/(N_{dat}-N_{par})}\big)$ is an acceptable representation of the surface thermophysical properties, and taking the standard deviation of ($\Gamma$, $f_c$, $\epsilon'$) in the set of acceptable solutions. Here, $N_{dat}$ is the number of independent data points, or the number of total pixels divided by the number of pixels per resolution element (see Section \ref{sec:data}).
\subsection{Modeling electrical properties}
\label{sec:electrical}
For a given (sub)surface temperature distribution, the intensity and polarization of thermal emission, and the variation of these properties with emission angle, are set by the dielectric constant and by the roughness and scattering properties of the surface. Broadly speaking, metal-rich surfaces have high dielectric constants, which lead to low emissivities and high polarization fractions. Apparent emissivity decreases from center to limb on the object due to the increasing emission angle, while polarization fraction increases. Surface roughness causes a given resolution element to encompass facets with a broader range of emission angles, which acts to erase the emission angle dependence of emissivity (but does not significantly affect its value at low emission angles), and to decrease the polarization fraction in part through cancellation of polarization vectors within a resolution element. Volume scattering can decrease both emissivity and polarization within limits, but requires a sufficiently long path length of radiation in the material. The properties of an object's thermal emission, in particular when spatially resolved, can thus constrain composition, roughness, and scattering properties. In the following sections we define terms (Section \ref{sec:defn}) and review the relevant dielectric properties of materials (Sections \ref{sec:dielectricproperties}) and models for particulates and mixtures (Section \ref{sec:dielectric_mixtures}). The expected emission and polarization behavior for different surface compositions and surface textures will be compared to that of Psyche in Section \ref{sec:results}. 
\subsubsection{Definitions} \label{sec:defn}
The complex dielectric constant $\tilde{\epsilon}=\epsilon'+i\epsilon''$ (also known as the permittivity) measures the polarizability of a material and is related to the complex refractive index $\tilde{n}=n+i\kappa$, where $n$ is the ratio of the speed of light in a vacuum to the speed of light in the material and $\kappa$ is the extinction coefficient, through the relation 
\begin{equation}
\tilde{\epsilon}=\tilde{n}^2 \label{eqn:n}
\end{equation}
which assumes a non-magnetic, non-conducting medium. From this the following relations follow: 
\begin{gather}
\epsilon'=Re(\tilde{n}^2) = n^2-\kappa^2\\
\epsilon''=Im(\tilde{n}^2) = 2n\kappa\\
n=\sqrt{\tfrac{1}{2}(|\tilde{\epsilon}|+\epsilon')}\\
\kappa=\sqrt{\tfrac{1}{2}(|\tilde{\epsilon}|-\epsilon')}\label{eqn:kappa}
\end{gather}
The emissivity and fractional polarization of a material as a function of emission angle $\theta$ are calculated from the Fresnel reflection coefficients for polarization parallel $R_{\parallel}$ and perpendicular $R_{\perp}$ to the direction of propagation:
\begin{gather}
E(\theta)=1-\tfrac{1}{2}[R_{\perp}(\theta)+R_{\parallel}(\theta)] \label{eqn:E}\\
F(\theta)=\frac{\tfrac{1}{2}[R_{\perp}(\theta)-R_{\parallel}(\theta)]}{E(\theta)} \label{eqn:F}
\end{gather}
with $R_{\parallel}$ and $R_{\perp}$ calculated from the Fresnel equations \citep{jackson1975}:
\begin{gather}
R_{\parallel}(\theta) = \bigg|\frac{\tilde{n}^2cos(\theta)-\sqrt{\tilde{n}^2-sin^2(\theta)}}{\tilde{n}^2cos(\theta)+\sqrt{\tilde{n}^2-sin^2(\theta)}}\bigg|^2 \label{eqn:Rpara}\\
R_{\perp}(\theta) = \bigg|\frac{cos(\theta)-\sqrt{\tilde{n}^2-sin^2(\theta)}}{cos(\theta)+\sqrt{\tilde{n}^2-sin^2(\theta)}}\bigg|^2 \label{eqn:Rperp}.
\end{gather}
Note that in the case where the magnetic permeability of the material $\tilde{\mu}$ differs substantially from that of free space, $\tilde{n}^2=\tilde{\mu}\tilde{\epsilon}$ instead of the relation given in Equation \ref{eqn:n}.\par
Emission at millimeter wavelengths arises from below the surface in most planetary materials; the electrical skin depth $\delta_{\textrm{elec}}=\lambda/(4\pi \kappa)$ defines the depth over which emission is reduced by 1/$e$ in a non-conducting material. Typically the loss tangent $\mathrm{tan \delta} = \epsilon''/\epsilon'$ is given instead of $\epsilon''$, and Equation \ref{eqn:kappa} then becomes
\begin{equation}
\kappa = \bigg[\frac{\epsilon'}{2}\Big(\sqrt{1+(\mathrm{tan \delta})^2}-1\Big)\bigg]^{1/2}.
\end{equation}\par
\begin{figure}[ht]
\centering
\includegraphics[width=12cm]{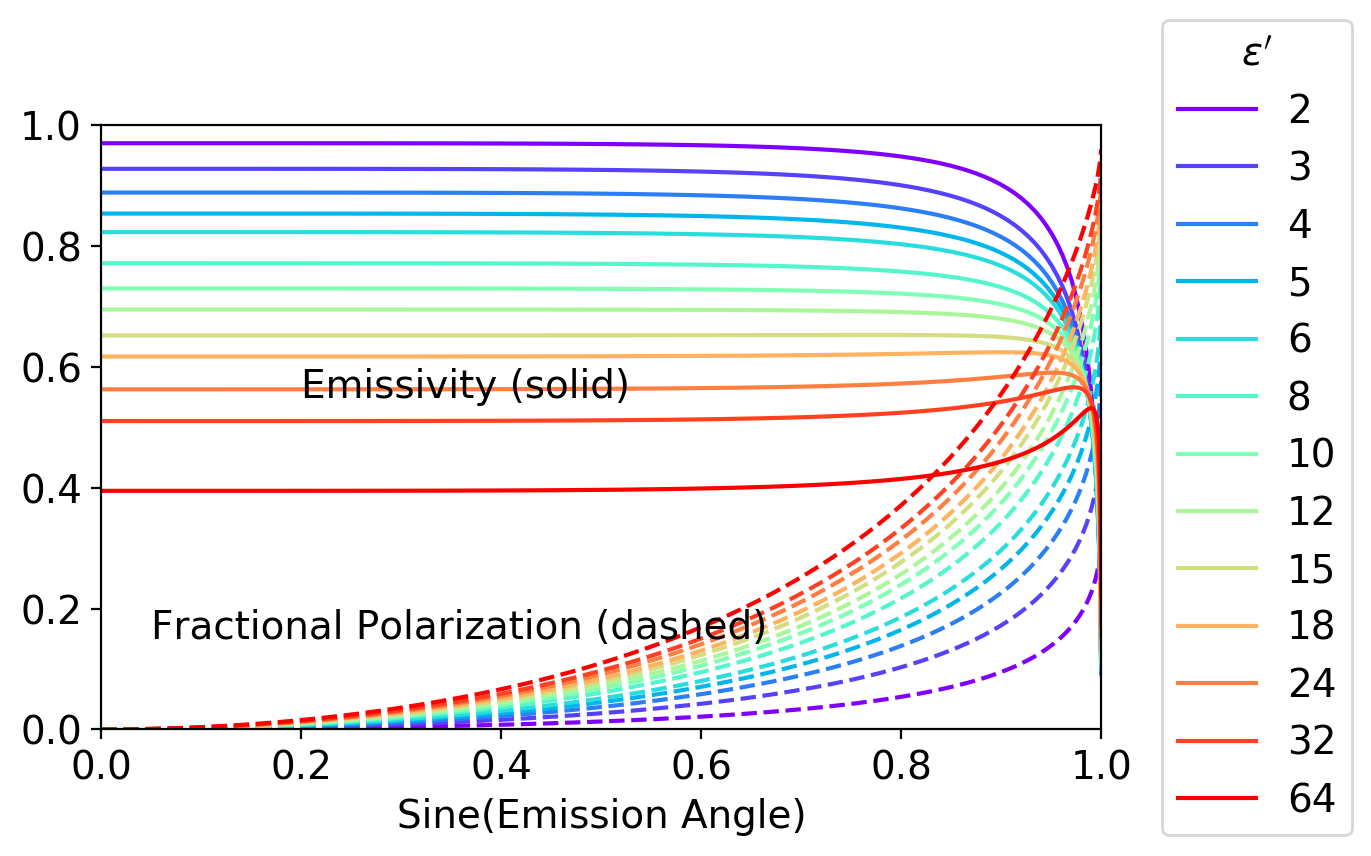}
\caption{Smooth-surface models for emissivity and fractional polarization for a range of values for the real part of the dielectric constant. \label{fig:smooth}}
\end{figure}

For a smooth surface and assuming that the imaginary part of the dielectric constant can be neglected (see Section \ref{sec:dielectricproperties}), the emission and polarization are shown in Figure \ref{fig:smooth} as function of emission angle for different values of $\epsilon'$, computed from Equations \ref{eqn:E} and \ref{eqn:F} . The plot is shown with the sine of the emission angle on the x-axis; this would be the fractional distance from disk center to limb for a spherical body, and is an approximate representation of that distance for an irregular ellipsoidal body such as an asteroid.\par
\begin{figure}[ht]
\centering
\includegraphics[width=14cm]{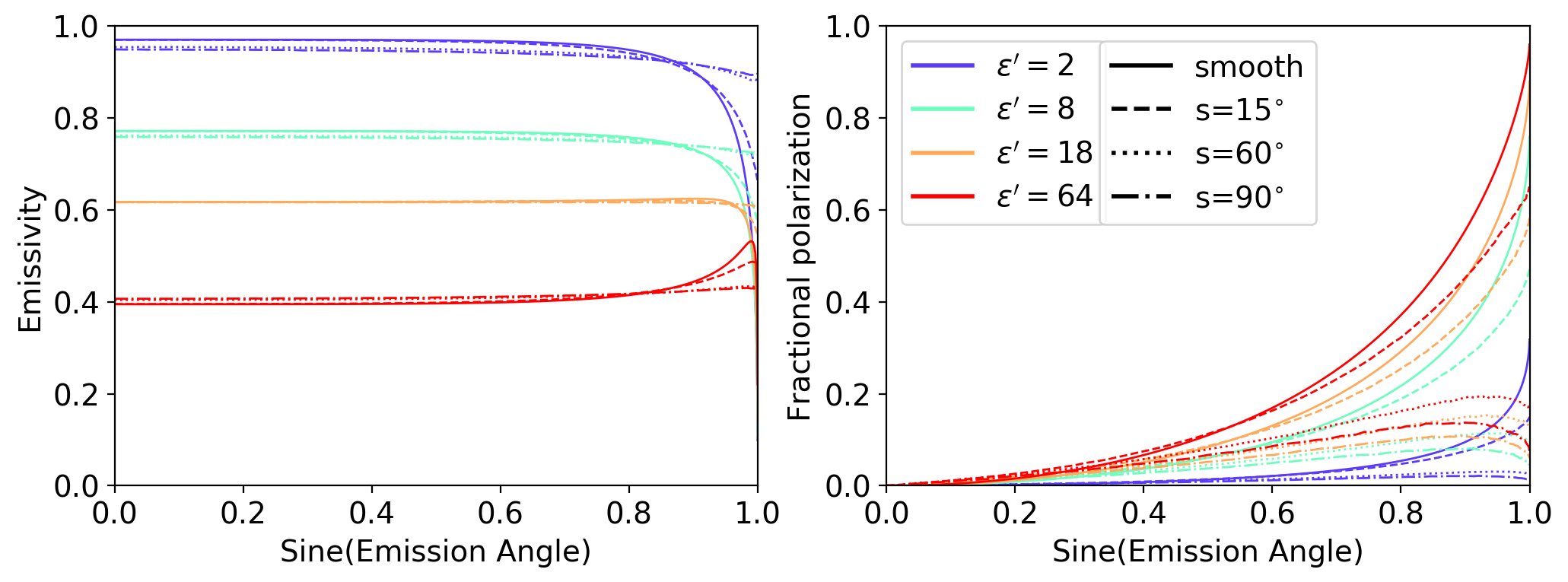}
\caption{Rough-surface models for emissivity and fractional polarization for a range of values for the real part of the dielectric constant and three values of the roughness parameter $s$, compared with the smooth surface models. \label{fig:rough}}
\end{figure}

Surface roughness can alter the profiles of both emission and polarization, in particular near the limb. We model the emissivity and polarization effects of surface roughness by treating the surface as a collection of facets far below the spatial scale of a resolution element but larger than a wavelength, and averaging the emission from a large collection of such facets. The emission intensity is scalar averaged, normalizing by the projected area of each facet, while the emission polarization is vector averaged to correctly treat the cancellation of the polarization vectors within a resolution element. Using facets that are randomly distributed in azimuthal orientation and gaussian distributed in angle from surface normal yields the models for $E(\theta)$ and $F(\theta)$ shown in Figure \ref{fig:rough}. The input parameter in the roughness model is the rms $s$ of the gaussian distribution from which the facet tilts are chosen. As can be seen in Figure \ref{fig:rough}, a higher surface roughness corresponds to a smaller change in both emissivity and polarization as emission angle increases, and hence erasure of the large polarization signature that is present near the limb in the smooth surface case. However, note that surface roughness has a minimal effect on the emissivity at low emission angles.
\subsubsection{Electrical properties of rocks, meteorites and metals} \label{sec:dielectricproperties}\label{sec:metalemission}
The dielectric constant of a material is in general a function of the frequency of radiation, and laboratory measurements of rocks and meteorites at millimeter wavelengths ($\sim$100-300 GHz) are very limited. However, numerous measurements exist at lower frequencies, and \cite{campbell1969} and \cite{hickson2020} find minimal variation in solid-rock permittivities across 20 MHz to 35 GHz. \cite{brouet2014} find that the real dielectric constant of particulate materials increases with increasing frequency from 50 MHz up to 190 GHz but stays within the 2-4 range. At grain sizes comparable to a wavelength they see evidence for volume scattering in the dielectric constant, but the effect on emissivity is below the 5\% level. We therefore use dielectric properties measured at lower frequencies than our observations without applying any correction, noting that this analysis would be improved by new lab measurements at ALMA-relevant frequencies.\par
Powdered rocks exhibit a value of $\epsilon'$ very close to 2 over a wide range of compositions, while solid rocks on Earth have values of $\epsilon'$ in the range of 2-10 \citep{parkhomenko1967,campbell1969,ulaby2014}. The lower end of the range is representative of porous materials such as pumice and tuff as well as high-silica rocks such as rhyolites. The high end of the range includes dense and highly mafic silicates; in general, higher metal content in a material translates directly to a higher dielectric constant. Individual minerals, such as metal sulfides and oxides, may have dielectric constants significantly higher than this. For example, $\epsilon'$ for FeS$_2$ and Fe$_3$O$_2$ is in the 34-81 range (measured at very low frequencies), and for Mn and Ti compounds is higher still \citep{parkhomenko1967}. \cite{campbell1969} also measured several chondritic meteorites at 450 MHz, and found $\epsilon'$ in the range of 8-12 for the L chondrites (7-11\% metal), 15-80 for the H chondrites (15-20\% metal), and 130-150 for the enstatite chondrite Indarch, which is 33\% iron by weight, primarily in the form of FeS (14\%) and metallic Fe (24\%) \citep{berthet2009}. \par
In a truly metallic, conducting surface, $\tilde{\epsilon}$ is purely imaginary and ohmic loss dominates over dielectric loss, in which case $\tilde{\epsilon}=i\sigma/{\omega}$ where $\sigma$ is the conductivity and $\omega=2\pi \nu$ for observing frequency $\nu$. Conductivities of pure metals are in the range of $10^6-10^7$ S/m, increasing with increasing temperature \citep{meaden1965}. For iron at 200 K, for example, the conductivity is roughly 7$\times$10$^6$ S/m. Such a material is almost completely reflective, with an emissivity $<<$1.\par
The meteorite measurements of \cite{campbell1969} found loss tangents in the range of 0.01--0.1, corresponding to $\kappa^2$ values of 0.14--0.33 for the three highest-loss samples and 0.0004 for the lowest-loss sample. This yields a $\kappa^2$/$\epsilon'$ ratio below 0.006 in all cases, and going forward we make the simplification $\tilde{n}^2 \sim \epsilon'$ when computing reflection coefficients from Equations \ref{eqn:E} and \ref{eqn:F}. The above extinction coefficients translate into values of $\delta_{elec}$ from a fraction of a wavelength up to a several wavelengths, and we adopt $\delta_{elec}$=2 mm in our emission calculations. The model fits are not highly sensitive to this parameter for the case of Psyche, where the subsurface temperature gradient is very shallow due to the high thermal inertia (to be presented in Section \ref{sec:results}). However, we note that future lab results on the loss tangents of relevant materials may motivate and enable more sophisticated treatment of the imaginary component.
\subsubsection{Electrical properties of particulates and mixtures} \label{sec:dielectric_mixtures}
For low-density particulate material, the real part of the dielectric constant is roughly set by the bulk density alone; the bulk density is in turn a function of the porosity and the solid grain density. Past studies have used this fact to convert radar reflectivity to bulk density and hence composition \citep{ostro1985,magri2001,shepard2010,hickson2020}. \cite{ostro1985} used the empirical formula R($\rho_{bulk}$)=0.12$\rho_{bulk}$-0.13 where R is the average Fresnel reflection coefficient, adopting densities of 3.2 and 7.8 g/cm$^3$ for rock and metal respectively, to jointly constrain the porosity and metal fraction of an asteroid regolith from the radar albedo. \cite{garvin1985} use the relation $\epsilon'(\rho_{bulk})=1.87^{\rho_{bulk}}$, which is an improvement on a formula derived from lunar measurements \citep{olhoeft1975}. \cite{hickson2020} evaluated a number of such formulas relating dielectric properties to composition and density, and instead find that the relationship $\epsilon'_{solid}=1.85\rho_{grain}+1$ for the solids, combined with the Looyenda-Landau-Lifshitz (LLL) model
\begin{equation}
(\epsilon'_{eff})^{1/3} = f(\epsilon'_{2})^{1/3}+(1-f)(\epsilon'_{1})^{1/3}
\label{eqn:LLL}
\end{equation}
to convert from solid to particulate (i.e. mixture of grains and free space), provided the best fit to their measurements and others in the literature. Here, $\epsilon'_{1}$ and $\epsilon'_{2}$ are the dielectric constants of the two materials, and $f$ is the volume fraction of material 2 in the mixture.\par
While models that derive dielectric constant from bulk density alone may provide a good fit to porous, low-metal-content regoliths, most are not able to produce the high dielectric constants of metal-rich materials. The linear relationship $\epsilon'_{solid}=1.85\rho_{grain}+1$ recommended by \cite{hickson2020} is unable to produce solid dielectric constants above 15, even for a grain density of 8 g/cm$^3$ (at the extreme upper end for meteorites), while the measured values for essentially all classes of chondrites are at this value or higher \citep{campbell1969}. \cite{ulaby1988} proposed a similar linear relationship between $\epsilon'_{solid}$ and $\rho_{grain}$, but added a correction term based on the content of various metal oxides. While this represents an improvement, their measurements were only made on silicate rocks and their linear models similarly cannot achieve $\epsilon'>15$. The power law $\epsilon'(\rho_{bulk})=1.87^{\rho_{bulk}}$ yields dielectric constants over 100, but only for densities approaching that of solid iron meteorites.\par
\begin{figure}[ht]
\centering
\includegraphics[width=8cm]{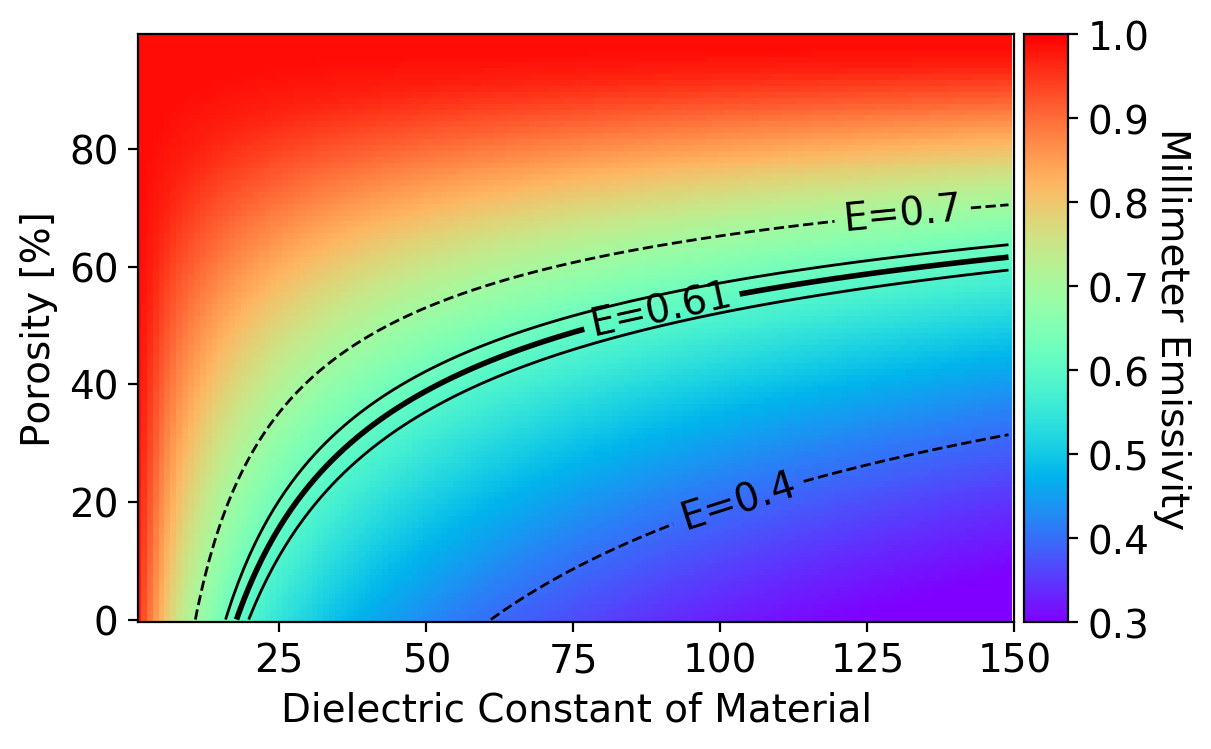}
\caption{Millimeter emissivity as a function of the dielectric constant of the solid material, and the porosity of the bulk material. The dashed lines are contours for the emissivities of 0.4 and 0.7, the bounds based on the zero and infinite thermal inertia cases, and three solid curves are contours for the best-fit value and its uncertainty, E$=$0.61$\pm$0.02, corresponding to a thermal inertia of 280.\label{fig:dielectricmodel}}
\end{figure}
\begin{figure}[ht]
\centering
\includegraphics[width=16cm]{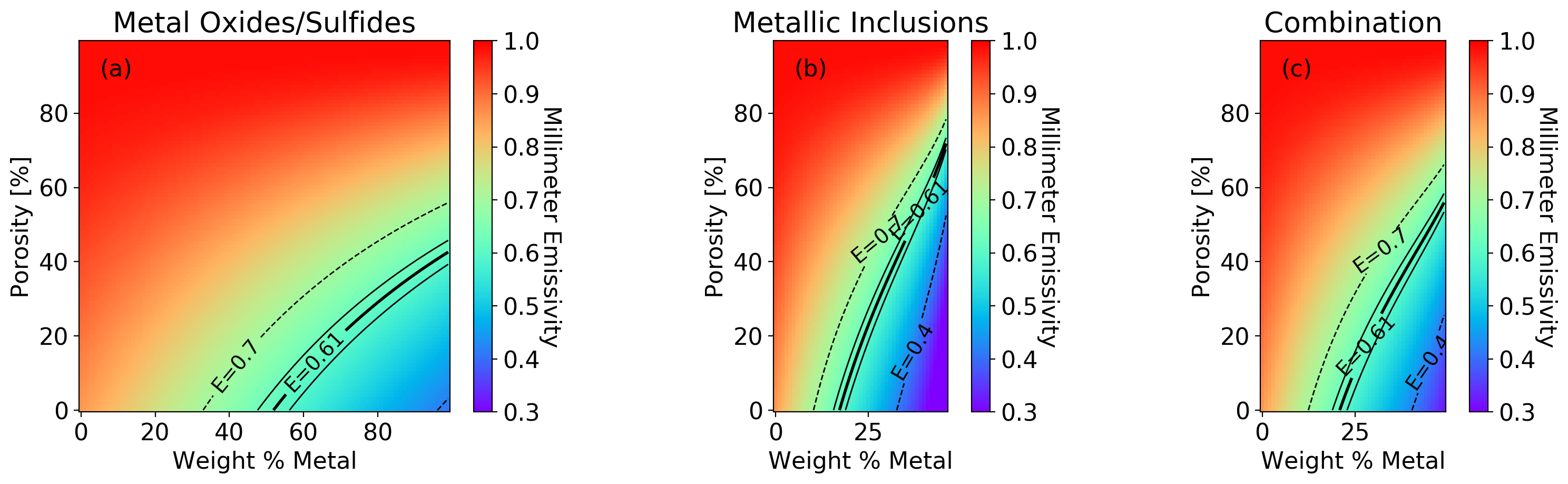}
\caption{Millimeter emissivity as a function of the porosity and weight percent metal of the bulk material, assuming (a) a dielectric mixture of $\epsilon'_{metal}=60$ and $\epsilon'_{rock}=5$ where the metal takes the form of oxides/sulfides, (b) conducting inclusions within a matrix with dielectric constant of $\epsilon'_{rock}=5$, and (c) a mixture in which 1/3 of the iron is in the form of sulfides and 2/3 is in the form of metallic inclusions, following the enstatite chondrite Indarch. The dashed lines are contours for the emissivities of 0.4 and 0.7, the bounds based on the zero and infinite thermal inertia cases, and three solid curves are contours for the best-fit value and its uncertainty, E$=$0.61$\pm$0.02, corresponding to a thermal inertia of 280.\label{fig:dielectricmodel_LLL_CI}}
\end{figure}

An alternate approach to estimating the dielectric constant of a mixture is to assume the dielectric constant of the constituent materials, but not to adopt any particular density dependence $\epsilon'(\rho)$. In the case of a particulate rock-metal mixture, Equation \ref{eqn:LLL} then becomes
\begin{equation}
(\epsilon'_{eff})^{1/3} = p(\epsilon'_{vacuum})^{1/3}+(1-p)f_{metal}(\epsilon'_{metal})^{1/3}+(1-p)(1-f_{metal})(\epsilon'_{rock})^{1/3}.
\label{eqn:LLL2}
\end{equation}
\indent Figure \ref{fig:dielectricmodel} shows the normal millimeter emissivity as a function of porosity and solid dielectric constant for a single-component porous material, calculated from Equations \ref{eqn:LLL} and \ref{eqn:E}. Figure \ref{fig:dielectricmodel_LLL_CI}a shows the millimeter emissivity of a porous mixture of rock and metal, calculated from Equation \ref{eqn:LLL2} and assuming values of $\epsilon'_{rock}=5$ and $\epsilon'_{metal}=60$, the latter of which is roughly in the middle of the range for iron sulfides and oxides measured by \cite{parkhomenko1967}. Densities of 3.2 and 4.6 g/cm$^3$ for rock and metal respectively are used to convert between volume fractions and weight fractions; the metal density is based on that of troilite and approximates the density of iron oxides and sulfides. This approach circumvents the need to adopt any empirical relationship between dielectric constant and density, but is heavily dependent on the assumed dielectric constants of the constituent materials. We note that this formula still underpredicts the dielectric constants of meteorites of 10-35\% metal content measured by \cite{campbell1969}; this ultimately arises from the fact that \cite{campbell1969} measured a higher dielectric constant for their enstatite chondrite meteorite than \cite{parkhomenko1967} did for pure metal oxides and sulfides. This suggests the presence of metallic inclusions, which can have a much stronger effect on the dielectric constant. \par
In particular, a metal-rock mixture may take the form of a ``loaded dielectric'', a model applied to Venus by \cite{pettengill1988,pettengill1992} that treats metal content in the form of conducting inclusions within a dielectric material, and calculates the $\epsilon'$ of the solid via
\begin{equation}
\epsilon'_{solid}=\epsilon'_{matrix}\frac{1+(1-\kappa')\gamma}{1-\kappa' \gamma}
\label{eqn:CI}
\end{equation}
where $\kappa'$ lies between zero and one and parameterizes the effective dielectric constant immediately surrounding the inclusions, and $\gamma$ relates to the shape of the particle. \cite{pettengill1988} use empirical values of $\kappa'=0.137$ and $\gamma=25.3f_{incl}$ where $f_{incl}$ is the volume fraction of conducting inclusions. These values are based on laboratory data from \cite{kelly1953}, who measured aluminum and pyrite fragments in a dielectric medium at a wavelength of 3.2 cm, and are calculated for $\epsilon'_{matrix}=5.0$. The results should be interpreted with caution given the mismatch in both inclusion composition and wavelength between the lab measurements and our observations, but the model has the qualitative advantage over the previously-discussed models of achieving high dielectric constants with metal content as low as tens of \%. This model is shown in Figure \ref{fig:dielectricmodel_LLL_CI}b, where $\epsilon'_{solid}$ from Equation \ref{eqn:CI} has been converted to the dielectric constant of the particulate via Equation \ref{eqn:LLL}. In this case, the metal density is taken to be 7.8 g/cm$^3$ in converting between volume and weight fraction because the metal is in pure metallic form rather than oxides/sulfides. The first two panels of Figure  \ref{fig:dielectricmodel_LLL_CI} together demonstrate that the conducting inclusion treatment of the metal component raises the dielectric constant for the same metal abundance, and hence allows for lower metal content and higher porosity for a given measured emissivity.\par
Finally, Figure \ref{fig:dielectricmodel_LLL_CI}c shows a combination loosely based on the Indarch EH4 enstatite chondrite meteorite, in which a rocky matrix of $\epsilon'=5.0$ contains metal in the form of both sulfides and conducting inclusions, the former making up 1/3 of the metal content and the latter making up 2/3 \citep{berthet2009}. The model is calculated using Equation \ref{eqn:LLL2} where the three components are air ($\epsilon'$=1), metal oxide/sulfide ($\epsilon'$=60), and rock containing conducting inclusions. The dielectric constant of the third component is computed from the conducting inclusion model described above. In this model, the ratio of metal content between oxides/sulfides and metallic inclusions is fixed at 1:2. Weight fractions are converted to volume fractions using densities of 3.2, 4.6, and 7.8 g/cm$^3$ for rock, oxides/sulfides, and conducting inclusions, respectively. \par
As discussed above, each of these methods has limitations, arising in large part from the fact that the dielectric properties of particulate metal-rock mixtures have not been measured at frequencies near 200 GHz. However, the existing measurements and parameterizations can provide a general framework for interpreting the observed millimeter emissivity of Psyche and comparing it to the measured millimeter emissivities of other asteroids. 
\section{Results and Interpretation}\label{sec:results}
\subsection{Thermophysical properties} \label{sec:TIresults}
For a thermophysical model with thermal inertia, surface roughness, and dielectric constant as the three free parameters, we find a best-fit thermal inertia of $\Gamma$=210$\pm$60, a smooth surface texture, and a dielectric constant of $\epsilon'$=21$\pm$1. The thermal and electrical properties are linked in the model: the surface roughness enters in both the thermal model and the emissivity profile, and the dielectric constant enters into both the emissivity and the subsurface integration calculations. However, the best-fit model does not provide a good fit to the variation in brightness with emission angle; residuals from even the best-fit models show a dark ring around the limb, indicating that the model is overpredicting either the temperature or the emissivity at high emission angles. Note that increasing surface roughness causes an increase in emission from high emission angles (Figure \ref{fig:rough}), and any surface roughness is therefore disfavored by the model.

\begin{figure}[ht]
\centering
\includegraphics[width=16cm]{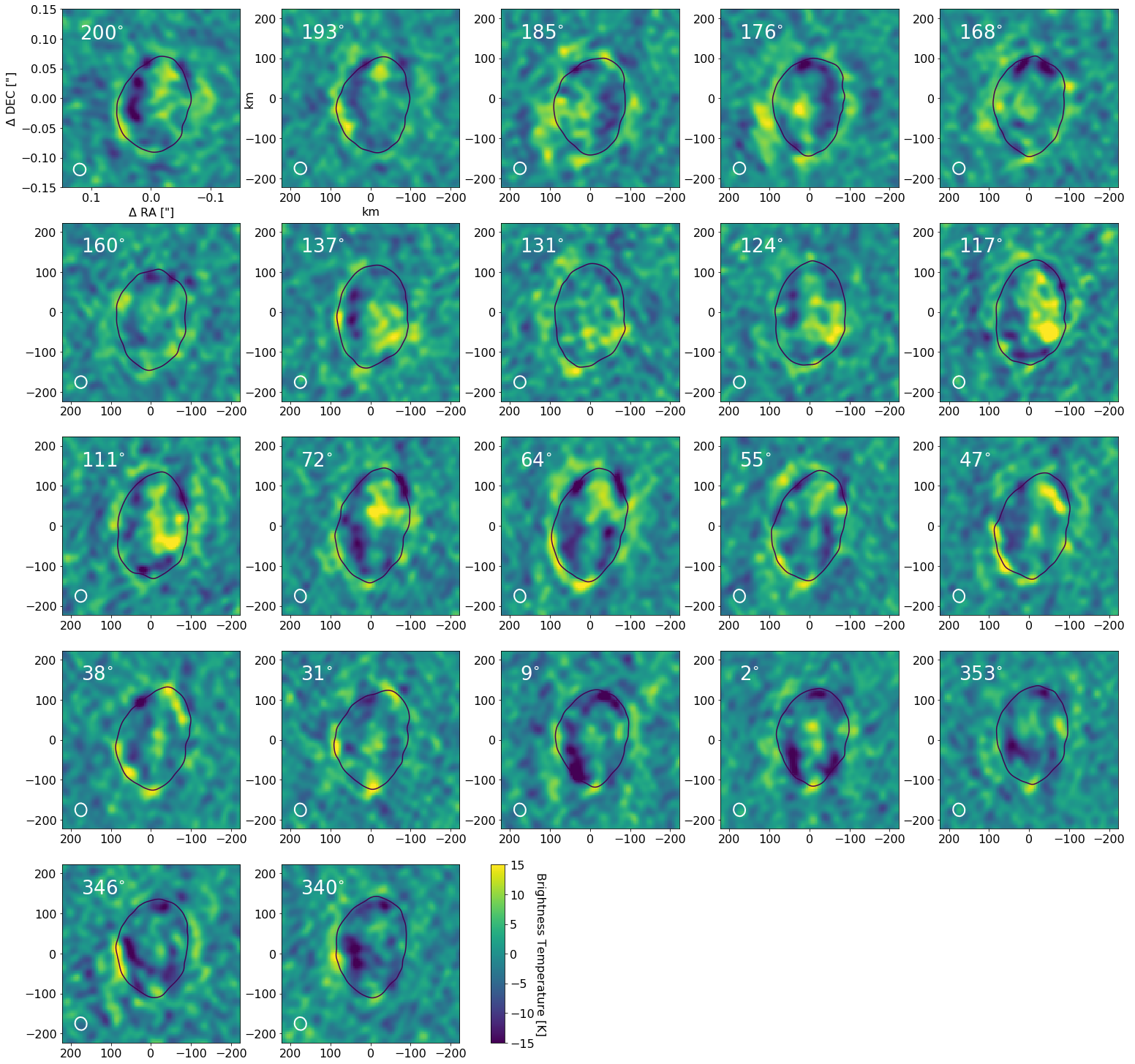}
\caption{Residuals from the images shown in Figure \ref{fig:ims} after subtraction of our best-fit thermal model, with thermal inertia of 283, no surface roughness (assumed), a dielectric constant of 18.5 parameterizing normal emissivity and a dielectric constant of 7 parameterizing the variation in emissivity with emission angle. The sub-observer longitude is indicated on each panel. The sub-solar and sub-observer latitudes are 3$^{\circ}$ and -14$^{\circ}$ respectively. An animation version of the images in this figure is provided with the paper.} \label{fig:resemi}
\end{figure}

Motivated by this shortcoming, which could only have been identified in spatially resolved data, we fix the surface roughness to that of a smooth surface based on the results of the above fits, but allow for the variation in emissivity with emission angle to be parameterized by a dielectric constant $\epsilon'_2$ that is different from that which sets the normal emissivity $\epsilon'_1$. In doing so, we explore the hypothesis that the dark rings are due to deviations of the asteroid surface from an ideal Fresnel surface. For a model with zero roughness and free parameters ($\Gamma$, $\epsilon'_1$, $\epsilon'_2$), we find a best-fit thermal inertia of $\Gamma$=280$\pm$100, a dielectric constant $\epsilon'_1$=18.5$\pm$0.4 and a dielectric constant $\epsilon'_2$=7.0$\pm$2.0. Including uncertainties on the flux density scale calibration, we find the larger allowed range of $\epsilon'_1$=17--21, corresponding to a normal emissivity of 0.61$\pm$0.02. This model provides a substantially better fit to the data, and resolves the problem of overestimation of emission at high emission angle. We therefore adopt this model as our preferred model and present its best-fit parameters as the preferred thermal inertia and dielectric constant in Section \ref{sec:smoothemission}-\ref{sec:conc}. The results will be discussed in the context of past thermal inertia measurements in Section \ref{sec:disc}. Residuals from the best-fit model are plotted in Figure \ref{fig:resemi}, where it can be seen that the residuals show no remaining systematic trends.\par
\subsection{Emissivity: Metal content and porosity} \label{sec:smoothemission}
We find a millimeter emissivity for Psyche of 0.61$\pm$0.02 based on our best-fit thermophysical model. We report a peak brightness temperature for Psyche of 114-129 K, where the range represents the peak value as a function of Psyche's rotation (Table \ref{tbl:obs}). For a non-rotating asteroid (STM) the peak temperature would be 239-265 K, while for the infinite thermal inertia (FRM) case it would be 180-185 K. These temperatures represent the absolute upper and lower bounds on possible peak surface temperatures, and would imply millimeter emissivities of 0.4 and 0.7 respectively. 

In the simplified case of a smooth, homogeneous, non-conducting surface material, the millimeter emissivity and fractional polarization, and their emission angle dependence, are completely determined by the dielectric constant of the material via the Fresnel equations as described in Section \ref{sec:defn}. The emissivity $E(\theta)$ and fractional polarization $F(\theta)$ for a set of smooth-surface models that differ only in dielectric constant are shown in Figure \ref{fig:smooth}. A higher dielectric constant translates directly to a lower emissivity, sharper emissivity drop at high emission angle, and higher degree of fractional polarization.\par
The measured emissivity of 0.61$\pm$0.02 (full possible range of 0.4-0.7) requires a value of $\epsilon'$ of 17-21 (full possible range: 12-60), which is higher than the typical range of Earth rocks ($\epsilon' \sim2-10$; see Section \ref{sec:dielectricproperties}), even including dense, mafic silicates. The bulk value of $\epsilon'$ is a function of both the solid grain dielectric constant and the porosity; a value of 20 can be matched by solid rock of this dielectric constant, or by a porous material with a higher grain dielectric constant (see Figure \ref{fig:dielectricmodel}). For example, a grain dielectric constant of 150 and a porosity of 60\% would also produce an emissivity of 0.60. For a model that treats the surface as a porous mixture of silicate rock and iron oxides/sulfides, the surface must be at least 50\% metal by weight with a porosity no greater than 40\% (see Figure \ref{fig:dielectricmodel_LLL_CI}a). If instead the metal takes the form of conducting metallic inclusions, the metal content may be as low as 20\% by weight if the material is completely solid (0\% porosity), or as high as 45\% (with 75\% porosity; see Figure \ref{fig:dielectricmodel_LLL_CI}b). However, while conducting inclusions may be present, it is unlikely that they are the sole metallic constituent. Analysis of the enstatite chondrite meteorite Indarch found that a third of its iron content was in the form of FeS rather than metallic Fe \citep{berthet2009}, and Psyche's near-infrared spectral slope is indicative of iron in the form of oxides \citep{gaffey1978}. If Psyche's iron content follows the ratio of FeS to metallic Fe in Indarch, the porosity cannot exceed 60\% (for the case of 50\% metal), and the metal fraction can be no smaller than 23\% in the no-porosity case (see Figure \ref{fig:dielectricmodel_LLL_CI}c). Note that all calculations assume no reduction in emissivity due to volume scattering (see Section \ref{sec:subsurfemission}).  \par
\subsection{Polarization}
\begin{figure}[ht]
\centering
\includegraphics[width=8cm]{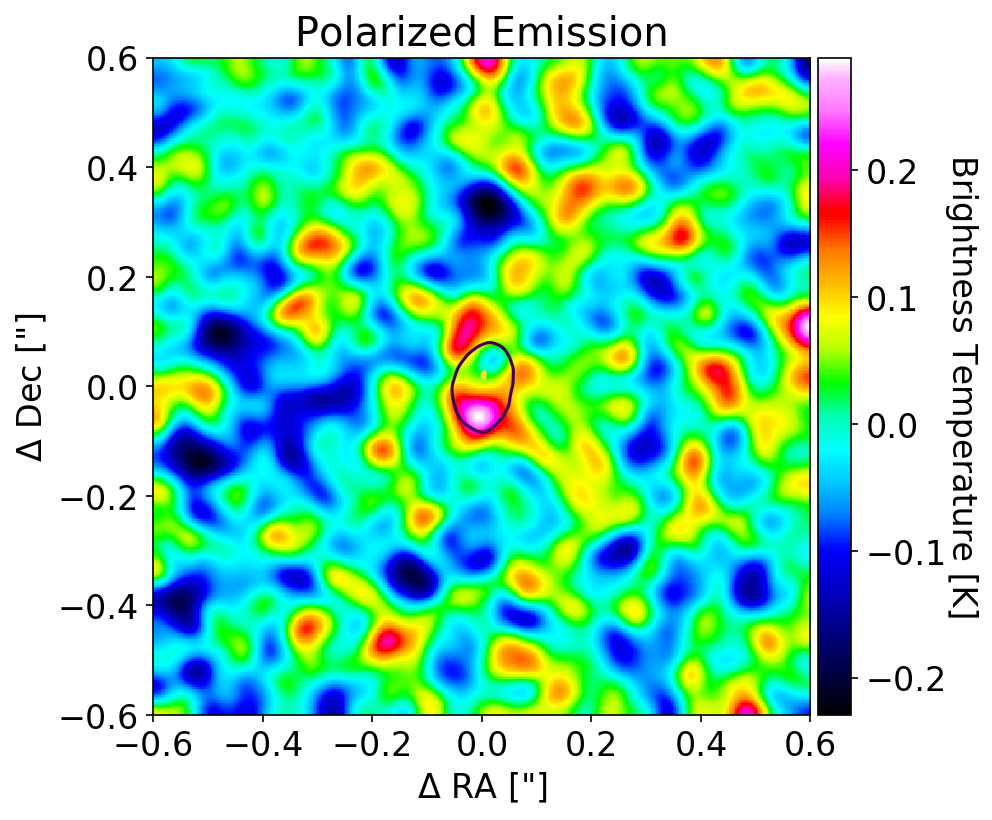}
\caption{Image of linearly polarized emission from Psyche, averaged across the full observing period and smoothed to lower resolution. The outline of Psyche's thermal continuum is shown in black, averaged across the images shown in Figure \ref{fig:ims}. The polarized emission arises predominantly from the southernmost part of the object, corresponding to Psyche's morning hemisphere, and is around 5$\sigma$ above the noise after smoothing to this resolution.\label{fig:polimage}}
\end{figure}

Psyche's millimeter thermal emission is nearly unpolarized. Figures \ref{fig:polimage} and \ref{fig:polfrac} show that the polarization fraction is only above the noise towards the limb and after averaging across the entire period of observation, reaching a peak of 0.010$\pm$0.002 at emission angles approaching 90$^{\circ}$. The polarized emission is also only detected from the ``lower half'' the asteroid, or the morning hemisphere (see Figure \ref{fig:polimage}). The significance of this is unclear: the morning hemisphere does not exhibit preferentially higher emission angles averaged over the rotation, nor is the polarization clearly attributable to a small subset of images or surface regions that might bias the distribution.\par
The very low polarization of Psyche's emission is  surprising: even Solar System bodies known to host rough, particulate surfaces exhibit polarized thermal emission at cm-mm wavelengths. For example, polarization of the Moon's emission was measured to be at the several \% level at 1.2 mm, despite a dielectric constant in the 2-3 range \citep{clegg1970}, and Mercury's limbward polarization is upwards of 10\% at cm wavelengths \citep{mitchell1994}. The high thermal inertia and high radar albedo of Psyche suggest a dense and metal-rich surface, which should lead to a higher polarization fraction relative to lunar-like regoliths while instead the opposite is observed. \cite{lagerros1996} also predicted that the polarization of asteroid thermal emission should be detectable even without spatially resolving the target. \par
The smooth surface models in Figure \ref{fig:smooth} show the expected fractional polarizations for different dielectric constants: the very low measured fractional polarization is completely inconsistent with the expected polarization for a dielectric constant in the range of 17-21 indicated by the emissivity. Moreover, the measured polarization is in fact inconsistent with any dielectric constant in this model, and a smooth, non-scattering surface is hence ruled out by the observations. In the following section, we explore potential surface properties such as surface roughness, scattering, and subsurface emission, and evaluate their ability to match the observed signatures.\par
\subsection{(Sub)surface properties}
\subsubsection{Surface roughness}
Broadly speaking, surface roughness acts to erase the change in emissivity with emission angle and erase the polarization of the emission, as shown in Figure \ref{fig:rough} based for the roughness treatment described in Section \ref{sec:defn}. These effects arise from the fact that a rough surface is made up of many facets with different local surface normal directions, so that the emission is averaged over a broad range of emission angles and the polarization vectors partially cancel. Although our modeling of Psyche's thermal emission strongly disfavors rough surface models (see Section \ref{sec:TIresults}), for completeness we briefly consider whether surface roughness could produce the observed polarization signatures. \par
The rough surface model has little effect on emissivity over most of the disk yet reduces fractional polarization, which qualitatively seems to match the characteristics of the data. However, even extremely rough surfaces of $s=90^{\circ}$ produce fractional polarizations $>$0.1 at the limb for the high dielectric constant needed to match Psyche's low emissivity. In addition, the fractional polarization in such cases increases gradually with projected distance from the object's center, whereas the data show no polarization except in the $\sim$10\% of the surface closest to the object's limb (see Figure \ref{fig:polfrac}). We therefore find that a rough surface alone is insufficient to match the observed characteristics.\par
\subsubsection{Subsurface scattering}
Scattering in the subsurface can also result in a decrease in both the emissivity and polarization fraction. The effects of scattering are maximized for wavelengths close to the average particle size of the regolith, for mixtures of materials with different dielectric behavior, and for low-loss media that allow long path lengths of radiation within the material. Scattering appears to be the only plausible mechanism for the very low polarization fractions observed on Psyche, and we propose that Psyche's surface particle composition and/or size distribution is strongly scattering of millimeter radiation.\par
Depolarization by scattering requires a surface with a sufficiently large electrical skin depth to allow for some scattering to occur, as well as particles that efficiently scatter millimeter radiation. Scattering is optimized for particle diameters within an order of magnitude of $\lambda /2\pi$ (i.e. circumference on the order of a wavelength), or roughly 20 $\mu$m - 2 mm diameter for our 1.3-mm observations. Indeed, Psyche's optical to near-infrared spectrum is best matched by a fine-grained metal in intimate association with an iron-poor silicate \citep{gaffey1978}, while the optical polarization signature implies a metal grain size of 20-50 $\mu$m \citep{dollfus1979}. However, the lunar regolith particles are also in this size range yet do not completely depolarize the 1.2 mm radiation, despite the lower intrinsic polarizability of the material \citep{clegg1970}. Highly reflective materials such as metals are particularly strongly scattering, and the polarization behavior of Psyche may arise from conducting inclusions within an iron-poor material that has a lower loss tangent and hence permits longer path lengths of radiation. This scenario is at least qualitatively consistent with the conducting inclusion model presented above as a possible explanation for Psyche's low emissivity. A key prediction of this scenario is that surfaces with lower emissivities would exhibit lower polarization fractions, exactly the opposite of the Fresnel case, because the scattering effects of highly reflective conductive particles in their regoliths dominate over the higher intrinsic polarizability of a metal-rich material.\par
Volume scattering has also been invoked to explain the low emissivities of numerous Solar System bodies, so it is worth exploring whether Psyche's low emissivity can also be explained by this process. The notably low emissivities of the icy galilean satellites have been attributed to multiple scattering in their surfaces, whose water ice compositions are highly transparent to mm-cm radiation, resulting in long path lengths that permit numerous scattering events \citep{muhleman1991}. In the lunar case, early work modeling Rayleigh scattering by scatterers of a single size found that scattering could produce a substantial (tens of K) decrease in the observed brightness temperature  \citep{england1975}, but it was later shown that a more realistic treatment of the scattering reduced the effect to a few K \citep{keihm1982}.
Similarly, while \cite{tryka1992} argue that an emissivity as low as 0.5 can be achieved by subsurface volume scattering (for application to the Venus highlands), \cite{wilt1992} show that this is not achievable with plausible materials. In particular, for scattering to lower the emissivity by this amount the radiation must be scattered off of voids or fully reflective inclusions with single scattering albedos around 0.95-1.0, and the matrix material must be effectively lossless. The required matrix loss tangents of 10$^{-3}$--10$^{-4}$ are lower than nearly all naturally occurring materials, except perhaps a very finely powdered pure silica \citep{wilt1992}. \par
The brightness temperatures of several other asteroids, most notably Vesta, have also been repeatedly measured to be anomalously low at (sub-)mm wavelengths, although nowhere near as low as the M-types \citep{redman1998}. Past work has attributed this to finely-powdered, highly porous regoliths that suppresses the emission via subsurface scattering \citep{redman1992,redman1998,mueller1998}. However, \cite{keihm2013} show that by accounting for subsurface emission and calculating temperatures from a thermophysical model the infrared through centimeter thermal emission from all four of the large asteroids Ceres, Vesta, Pallas, and Hygeia can be well matched by a particulate surface with a low thermal inertia, an emissivity of 0.95, and self-heating due to roughness, with no need to invoke subsurface scattering.\par
Thus it has been repeatedly shown for many different targets that the extent of emission reduction due to subsurface scattering  is very minor for plausible surface conditions, even for porous, transparent surfaces. The extent of the effect is further reduced for an object such as Psyche that is expected to have a high bulk loss tangent, which does not permit a sufficiently long path length of radiation in the material for multiple scattering to effectively suppress emission. Indeed, while \cite{brouet2014} found evidence for volume scattering in their 190 GHz measurements of powdered volcanic rocks and ash, the effect corresponds to a $<5\%$ difference in emissivity and moreover was not present in the samples with high iron oxide content, where they infer that the dielectric losses dominate over scattering losses. We therefore conclude that volume scattering, while a plausible explanation for the depolarization, is not able to suppress emission from Psyche at a significant level. \par
\subsubsection{Subsurface emission} \label{sec:subsurfemission}
\cite{redman1998} explored the effects of subsurface emission on inferred emissivities, noting that the effective emissivities (defined as emission relative to that of a non-rotating spherical blackbody) should drop from roughly 1 at infrared wavelengths to lower values in the radio due to emission from the colder subsurface. As discussed in the previous section, this explanation was further quantitatively developed by \cite{keihm2013}, who showed that the previously-inferred low emissivities of asteroids including Vesta could be fully attributed to emission from the cold subsurface. Spatially-resolved observations of the asteroid Juno at 1.3 mm also found an emissivity of 0.8 relative to the modeled surface temperature, which was attributed to emission from the cooler subsurface \citep{junoalma}. The low emissivities of cold icy Solar System bodies have similarly been attributed to subsurface emission, and in addition to scattering as described previously \citep[ e.g.][]{trumbo2018,dekleer2021,muhleman1991,lellouch2017}.\par
The major difference between these examples and that considered here is the small electrical skin depth and high thermal inertia of Psyche. Within the bounds where subsurface emission plays a role, a low-density regolith and lossless surface materials contribute to lower apparent emissivities, because emission arises from deeper within surface for a more transparent material and because the subsurface temperature gradient is steeper for lower thermal inertias. In the case of Vesta \citep{keihm2013}, the extreme subsurface temperature gradient of 50 K/mm, combined with $\delta_{elec}$ on the order of millimeters, led to a severe underestimate of surface temperature when subsurface emission was not accounted for. In contrast, for Psyche's thermal inertia of 200-300, the temperature gradient is close to 1 K/mm, and is still as small as $\sim$3-5 K/mm for the somewhat lower thermal inertia of 100-150 found by \cite{matter2013}, while $\delta_{elec}$ is likely on the order of 1-2 mm or less (see Section \ref{sec:dielectricproperties}). This indicates that any possible underaccounting of subsurface emission in our model is not a plausible explanation for Psyche's low emissivity.\par
The argument above relies in part on the finding that Psyche's thermal inertia is high and hence the thermal gradient is shallow and the diurnal thermal skin depth larger than the electrical skin depth. However, two discrepant findings on Psyche's thermal inertia exist in the literature \citep{matter2013,landsman2018}, and in the lower thermal inertia case \citep[$\Gamma=5-25$, ][]{landsman2018} the temperature gradient is steep and the diurnal thermal skin depth much smaller than an electrical skin depth such that with ALMA we would be seeing emission from below the depth of diurnal temperature variations. This scenario is strongly disfavored by our data, but we note that even below the diurnal thermal skin depth the temperature can be no lower than 180 K (Section \ref{sec:thermal}), so even in such a case the data constrain the emissivity to be no higher than 0.70. 
\section{Discussion}\label{sec:disc}
\par
\subsection{Psyche's thermal inertia in context}\label{sec:discTI}
\begin{figure}[ht]
\centering
\includegraphics[width=16cm]{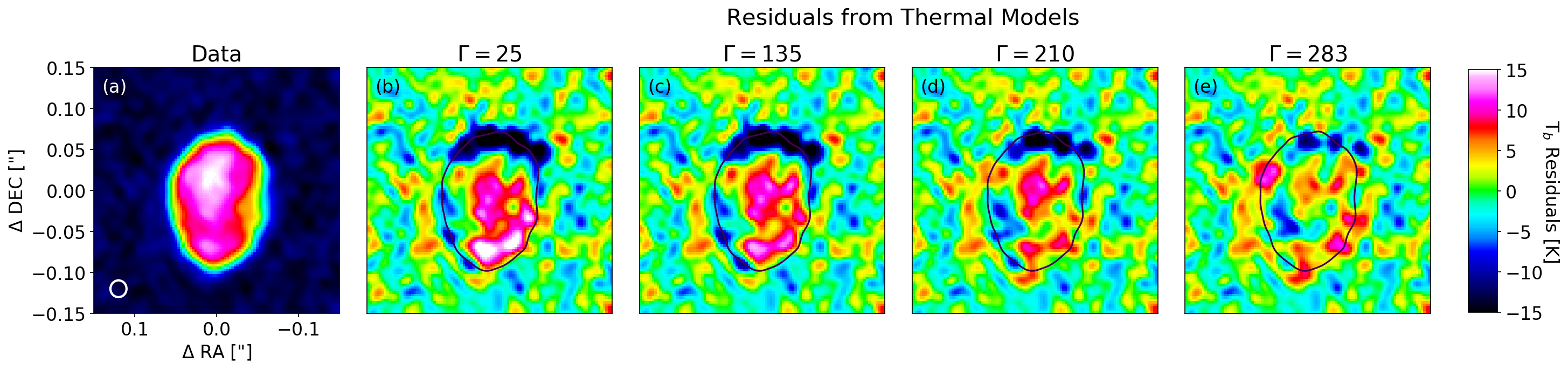}
\caption{Residuals from thermal models, shown for the single example image in (a). (b) Residuals from a thermal model using a thermal inertia of $\Gamma$=25, in the range found by \cite{landsman2018}; (c) using a thermal inertia of $\Gamma$=135, in the range found by \cite{matter2013}; (d) our best-fit model given the free parameters ($\Gamma$,$f_c$,$\epsilon'$), in which $E(\theta)$ is set by $\epsilon'$ following the Fresnel equations; and (e) our preferred best-fit model, with a smooth surface and free parameters ($\Gamma$,$\epsilon'_1$,$\epsilon'_2$) such that the shape of $E(\theta)$ is allowed to vary independently from the disk-center emissivity (i.e. not requiring a strictly Fresnel surface). \label{fig:TIex}}
\end{figure}

Two discrepant values exist in the literature for Psyche's thermal inertia. \cite{matter2013} found a thermal inertia of 133$\pm$40 or 116$\pm$40 for smooth and low-roughness models respectively, while \cite{landsman2018} found a best-fit value of 5-25 and a smooth surface. Our best-fit model has a smooth surface texture and a thermal inertia of 280$\pm$100. This is consistent at 1$\sigma$ with the smooth-surface value of \cite{matter2013}, but completely inconsistent with the lower value found by \cite{landsman2018}. Residuals from an example ALMA image for these various model cases are shown in Figure \ref{fig:TIex}. The figure shows a clear morning-to-afternoon temperature gradient in the residuals for the low thermal inertia models, illustrating that such models are not a good fit.\par 
While the thermal inertia values derived from millimeter and infrared observations of Solar System objects are often quite different, the very small electrical skin depth of a surface with high metal content means that for an object like Psyche the millimeter emission should arise from the very shallow subsurface (for Psyche, $\delta_{\textrm{elec}}\sim$ 1--2 mm, see Section \ref{sec:subsurfemission}), and shallower than a diurnal thermal skin depth ($\delta_{th}\sim$ 10--12 mm; see Section \ref{sec:TPM}). The infrared thermal inertia is therefore likely to still be a good approximation for Psyche at millimeter wavelengths. However, it is important to note that our model differs from those used to fit past mid-infrared observations. Those differences include treating emission from the subsurface (as is needed for millimeter observations); fitting well resolved data; and parameterizing the variation in emissivity with the emission angle based on the electrical properties of the surface. Finally, the introduction of an additional model parameter to decouple the emissivity from the shape of its emission angle variation was needed to achieve a good fit, but would not have been well-motivated without the spatial resolution to identify the dark limb rings in the residual images. This highlights the power of spatially resolved data, but also the difficulty in comparing model results between resolved and unresolved observations.
\subsection{Psyche's emissivity in context}\label{sec:context}
Most past observations of asteroids at millimeter wavelengths have not been disk resolved nor measured polarization. Typical asteroid regoliths exhibit disk-integrated emissivities between 0.9 and 1.0 and, as noted earlier, lower reported values have typically been explainable by subsurface emission from insulating regolith layers \citep{keihm2013,junoalma}. The most detailed measurements at these wavelengths have been made by the Microwave Instrument for Rosetta Orbiter (MIRO) on the \textit{Rosetta} spacecraft, which observed the asteroids \v Steins and Lutetia at wavelengths of 0.53 and 1.6 mm. A (sub-)mm emissivity value of $>$0.9 was determined for Lutetia, which appears to host a very insulating surface layer overlying a higher thermal inertia layer, while an emissivity of 0.6-0.7 at 0.53 mm and 0.85-0.9 at 1.62 mm was found for the rocky object \v Steins \citep{gulkis2010,gulkis2012}. The authors note that the emissivity of \v Steins at 0.53 mm was not well constrained and could in fact be as high as 0.8. \par
The emissivities reported for the M-type asteroids have always been unusually low. Out of seven asteroids observed from 0.45 to 2.0 mm by \cite{redman1998}, only the M-types Psyche and Kleopatra had emissivities too low even to be consistent with the deep surface temperatures. The authors interpreted this as evidence that the surfaces are reflective rather than emissive and attributed this to their metal-rich compositions. Their observations of Psyche at a wavelength of 1.3 mm found a flux density of 74$\pm$10 mJy \citep{redman1998}. Scaling this measurement from the solar and Earth distances at the time of their observations ($D_{sun}$=2.74 AU; $D_{Earth}$=1.74 AU) to the time of ours yields a flux density of 53$\pm$7 mJy, consistent with our measurements. While their inferred emissivity of $\sim$0.3 disagrees with ours, their emissivity was calculated from the disk-integrated flux density assuming a spherical, nonrotating blackbody using the size estimate from IRAS of D$=$254 km. Given the equatorial view at their time of observation, Psyche's true dimensions of $234\times 171$ km yield an area only 62\% of that assumed by \cite{redman1998}, raising their measured emissivity to $\sim$0.5 relative to a nonrotating blackbody. This is within the range we find from our data for the STM (nonrotating, no subsurface emission) model. \par
Our measured emissivity of 0.61$\pm$0.02 is calculated from our thermophysical model. Values of lower thermal inertia correspond to higher model surface temperatures and hence lower emissivity. For example, a thermal inertia of 100-150 corresponds to an emissivity in the 0.55-0.60 range. A comparison of measured and modeled disk-integrated flux densities still finds comparable emissivity values, indicating that at least for emissivity, results from resolved and unresolved datasets are directly comparable. \par
The unusually low emissivity found for Psyche is consistent with the high thermal inertia found by our models and by \cite{matter2013}; a denser surface with higher metal content will also have a higher thermal inertia. However, it would be challenging to reconcile with the fine-grained silicate regolith scenario proposed by \cite{landsman2018}. If Psyche's thermal inertia were in the 5-25 range \citep{landsman2018}, the diurnal surface temperature variations, as well as the actual surface temperatures, would be considerably higher than observed in our data and our derived emissivity would be close to 0.4. The discrepancy is mitigated somewhat by the fact that the thermal gradient is much steeper in such a case, and our observations would likely be seeing emission from below a diurnal thermal skin depth. However, even considering this effect, the emissivity of the subsurface material could be no higher than 0.70, which cannot be matched by a silicate regolith. A scenario in which an insulating layer overlays a dense, metal-rich layer could perhaps reconcile these results, but the insulating layer would need to be extremely thin ($<<1$ mm) for our data to be effectively insensitive to it.\par
Finally, in the case of a truly metallic surface, the surface material would be almost completely reflective, with an emissivity orders of magnitude below 1 (Section \ref{sec:metalemission}), and this scenario is easily ruled out for any porosity. However, if half of Psyche's surface were conductive, and the conductive and dielectric regions were interspersed below the resolution of ALMA (but with the regions much larger than a wavelength), the observations could be matched by a bulk surface that is lossy and scattering (e.g. a porous, non-metallic regolith), spatially interspersed with metallic regions that are effectively non-emitting. This idealized scenario requires the conducting regions to be pure, unweathered metal, which is implausible for an exposed and space-weathered surface although the space weathering of metals is not well understood. When the metal takes the form of oxides and sulfides, the real dielectric constant falls in the 20-150 range measured by \cite{parkhomenko1967} and \cite{campbell1969}. A surface composed of pure iron oxides/sulfides is consistent with Psyche's emissivity for a surface porosity of $\sim$40\% (Figure \ref{fig:dielectricmodel_LLL_CI}a), but the polarization signature would be challenging to explain in such a scenario. \par
\subsection{Rotational variability}\label{sec:variability}
The temperatures across Psyche's disk are relatively uniform, only varying within a $\sim$10 K range from the morning to evening terminators in most cases. This contrasts sharply with the S-type asteroid (3) Juno, for which variations of 50 K were observed \citep{junoalma}. The temperature variations across Psyche can be seen in Figures \ref{fig:ims} and \ref{fig:resemi}; the latter shows the residuals from our best-fit thermophysical model. The model includes a decrease in emissivity with emission angle following the functional form of Equation \ref{eqn:E} for a dielectric constant of 7.0, which was needed to match the emissivity at high emission angle. Note that this dielectric constant is only used to parameterize the variation in emissivity with emission angle, and differs from the higher dielectric constant required to match the emissivity. This difference could arise from scattering, and is not surprising given the lack of polarization, which already indicates that Psyche's surface is not well modeled by a smooth dielectric interface.\par
Over the fraction of Psyche's rotation covered by our dataset, the disk-averaged temperature varies from 88.0-97.7 K and the peak temperature varies from 115.4 to 128.9 K. These variations may arise from spatial variations in surface properties such as albedo and thermal inertia, which would correspond to actual variations in surface temperature, or from variations in millimeter emissivity. The ratio of maximum to minimum temperature is roughly 1.12 when considering either the disk-averaged temperatures or the peak temperatures of each scan. Because $T\sim(1-A)^{1/4}$, the albedo variations required to match these temperature variations would be very large; for example, a minimum Bond albedo of 0.02 and a maximum Bond albedo of 0.35. In the absence of phase integral variations across Psyche, albedo variations at this level are inconsistent with Psyche's visible light curve. If the variability arises solely from emissivity differences, variations at only the 12\% level are sufficient because $T\sim E$. When the best-fit emissivity is computed for each observation individually, the emissivity varies from 0.59 to 0.63, corresponding to $\epsilon'=17-21$.\par
\begin{figure}[ht]
\centering
\includegraphics[width=16cm]{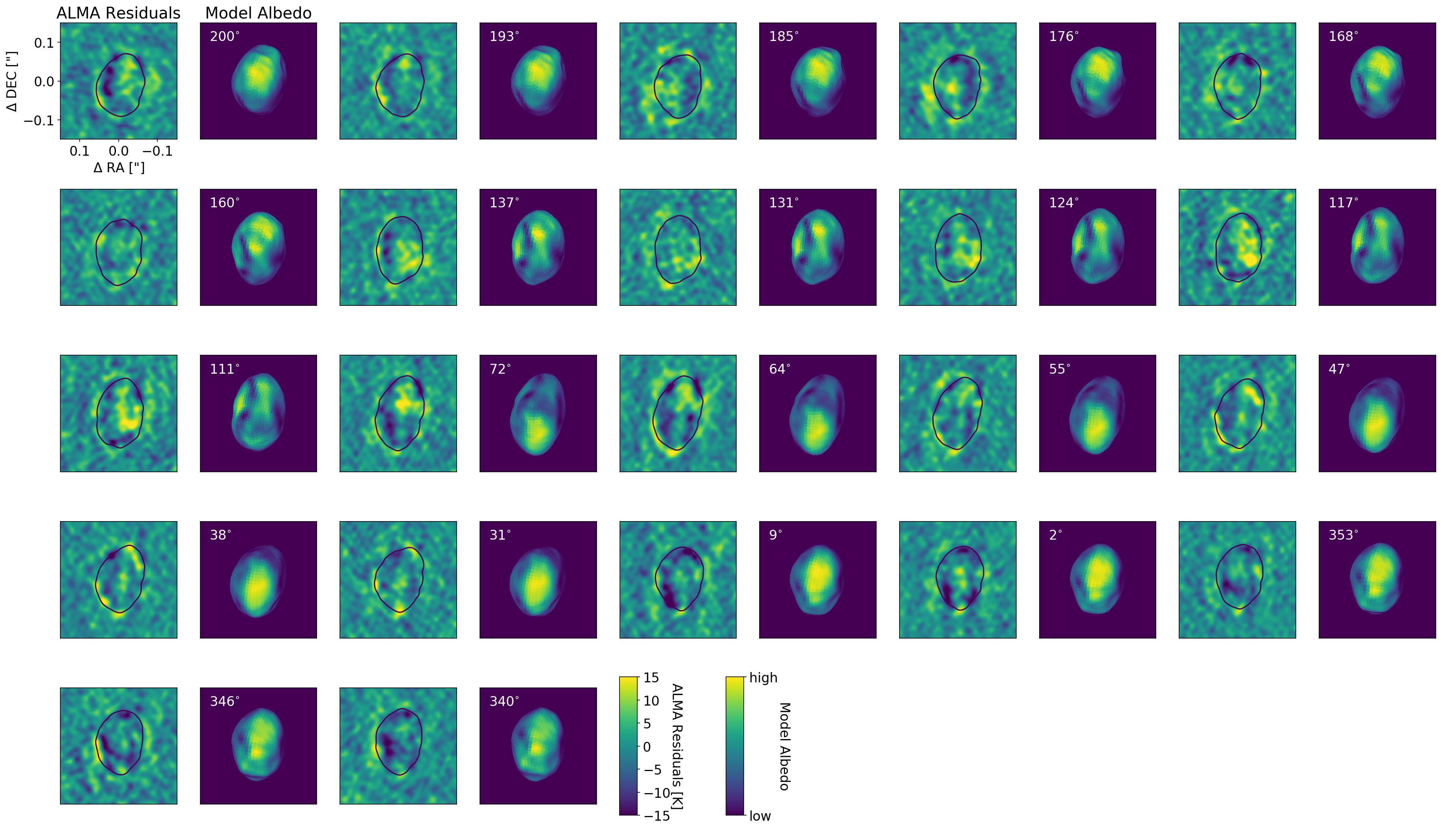}
\caption{Same as Figure \ref{fig:resemi} but with the projected albedo map from \cite{ferrais2020} interspersed at the viewing geometry of each observation. The albedo variations are exaggerated to highlight trends. Albedo and residuals should be anti-correlated if albedo alone is responsible for deviations of surface temperature from the best-fit model. \label{fig:resemi_wA}}
\end{figure}

 
The spatial resolution of the data may allow these global variations to be linked to specific locations on Psyche's surface. Variations in both visible and radar albedo across Psyche's surface are also observed \citep{ferrais2020,shepard2021} and suggest localized regions with distinct surface properties or enhanced metal content. A detailed correlation of datasets is beyond the scope of this paper, but Figure \ref{fig:resemi_wA} shows the \cite{ferrais2020} albedo map projected to the same viewing geometry as each of our observations. Datasets of this type, where individual surface regions are resolved and their temperatures tracked over the day, provide a powerful tool for detecting spatial variations in surface properties and mapping them back to the underlying variations in composition. \par
\subsection{Comparison with radar}\label{sec:radar}
Kirchoff's law for thermal radiation states that, for a body in thermal equilibrium, the amounts of energy absorbed and emitted are equal. This law is stated in terms of emissivity, $E$, and reflectivity, $R$ in Equation \ref{eqn:E} and is abbreviated as simply $E=1-R$. Assuming the regolith dielectric constant does not vary significantly between mm and cm wavelengths \citep{campbell1969,hickson2020}, we may use this relationship to check for consistency between the radar albedo at S-band (12.6 cm) and emissivity at 1.3 mm.\par
\cite{magri1994,magri2001} describe an approximate relationship between an asteroid's radar albedo, $\hat {\sigma}_{oc}$, and surface Fresnel reflectivity, $R$, for asteroids with radar echoes dominated by first surface reflections:
\begin{equation}\label{gain}
\hat {\sigma}_{oc} = gR 
\end{equation}
where $g \ge 1$ is the gain of the asteroid, an enhancement factor for reflected incoming energy. For a smooth sphere, $g=1$, while for a roughened asteroid surface the gain may be a few tens of percent higher than this \citep{magri2001}. \cite{shepard2015,shepard2017,shepard2021} report that Psyche's radar echoes are consistent with a relatively smooth surface and justify the application of this approximation. \cite{shepard2021} report Psyche's overall S-band radar albedo to be $\hat {\sigma}_{oc} = 0.34 \pm 0.08$, with maximum observed values as high as $\hat {\sigma}_{oc} = 0.52$. Given our assumptions and uncertainties, we adopt the first order estimate of Psyche's surface/near-surface Fresnel reflectivity at $\lambda = 12.6$ cm to be $R \approx 0.3 - 0.5$, and hence emissivity $E\approx 0.5 - 0.7$. This is consistent with our estimate of emissivity at $\lambda = 1.3$ mm and suggests that the regolith is similar in composition and bulk density from depths of a few millimeters to tens of centimeters.\par
Past observations of reflectivity and polarization properties that remain constant across radar wavelengths have been used to infer a power law distribution of scatterer sizes \citep{ostro1992,black2001}. The match between the 1.3 mm and 12.6 cm reflection  supports our assumption that volume scattering is not responsible suppressing Psyche's emissivity, since volume scattering by particles in a specific size range would be needed to effectively suppress emission, and such a particle size distribution would have a strongly wavelength-dependent effect. \par
\section{Conclusions} \label{sec:conc}

Psyche's thermal inertia of 280$\pm$100 is higher than typical values for large asteroids, consistent with a surface of higher density and thermal conductivity. However, it still falls far short of rock-like values (e.g. the 450-850 range found for \v{S}teins; \citet{gulkis2010}), indicative of a porous or regolith-covered surface.\par
Psyche's emissivity of 0.61$\pm$0.02 requires a bulk surface real dielectric constant of 17--21, higher than essentially all Earth rocks and significantly higher than the regolith observed on the terrestrial planets or the Moon. Considering models that treat the surface as either a dielectric mixture of silicates and iron oxide/sulfides, or as a low-loss material with metallic inclusions, the metal content must be at least 20\% (for the case of 0\% porosity). Higher metal contents are permitted at higher porosities, but a fully metallic, conducting surface of any porosity is ruled out. \par
Psyche's thermal emission is almost completely unpolarized, which is surprising given that even rough, insulating regoliths in the Solar System such as those of the Moon, Mars, and Mercury, exhibit polarized thermal emission at mm-cm wavelengths, and denser, more metal-rich surfaces should exhibit higher polarization fractions. While surface roughness lowers the polarization of thermal emission, it is unable to lower it as completely as is required to match the observations given the high dielectric constant implied by Psyche's emissivity, and is moreover strongly disfavored by our thermophysical model.\par
Scattering appears to be the only plausible mechanism to remove the polarization of the emission and suggests the presence of reflective metallic particles in the size range of tens of microns to millimeters. This is consistent with past optical observations, from which metal particle sizes of 20-50 $\mu$m were inferred \citep{gaffey1976,chapman1973} and with the low radar circular polarization ratio (SC/OC=0.1$\pm$0.1; \cite{shepard2008,shepard2017,shepard2021}), as such particle sizes are not strongly scattering of 12.6-cm radar waves. The consistency between the millimeter emissivity and the S-band radar reflectivity supports the conclusion that Psyche's surface properties do not vary over the upper tens of cm in the surface. A scattering surface is also qualitatively consistent with the fact that a simple Fresnel surface does not provide a good fit to the center-to-limb emission profile of Psyche observed by ALMA, and a lower dielectric constant is required to match the observed emission at high emission angle.  A porous, low-dielectric surface containing conducting inclusions tens to hundreds of $\mu$m in size is thus consistent with several results from optical through radar wavelengths.\par
Objects composed of high dielectric materials should exhibit lower emissivities and rapidly increasing fractional polarization towards the limb, as seen in Figure \ref{fig:smooth}. A surprising prediction of the metallic inclusion (aka ``loaded dielectric'') model is that, while their emissivities are consistent with their higher bulk dielectric constant, their fractional polarization will not increase with emission angle as rapidly as expected because volume scattering randomizes it and may erase the signature altogether. Future detailed modeling of the emission angle dependence of the emission and polarization presented here has the potential to yield new insight into the grain properties and size distribution. \par
We compare the properties of several candidate compositions for Psyche's surface against the emissivity and polarization properties in Table \ref{tbl:scenarios}. An enstatite chondrite composition has the desired properties of a relatively low-loss silicate matrix containing substantial metal, and the porosity required to match the emissivity yields a surface density of 3.2 g/cm$^3$, consistent with the results of past work. CB chondrite,  pallasite, and mesosiderite compositions also cannot be ruled out, although Psyche's polarization properties are difficult to explain if the constituent silicates are olivine or pyroxene. The emissivity can also be matched by a scenario in which solidified ferrovolcanic flows composing $\sim$40\% of the surface area are interspersed with regions of low-Fe silicates below the resolution of our data, but the polarization properties are again difficult to explain in such a case. \par
Only a few asteroids of this size have had their millimeter emission measured and spatially resolved, and none with polarization measurements. As such, it is challenging to put these observations in context until a greater number of objects have been studied in this way. Future observations over a larger range of asteroid classes will provide new perspective into how regolith properties correlate with composition, and may reveal variability in regolith properties within asteroid classes.\par
Observations over a wider frequency range would be particularly valuable. If the depolarization presented here arises from a size distribution of reflective particles tens to hundreds of $\mu$m in size, then thermal emission at mid-IR or cm wavelengths should show a stronger polarization signature. Radar measurements ($\sim$10 cm wavelength) suggest that this is the case, but passive thermal measurements sensitive to the same subsurface depths and particle sizes as radar would provide a useful reference point. The models described here to interpret the low emissivity are heavily limited by the lack of measurements of the dielectric properties of metal-rich materials in both solid and particulate form, near 200 GHz. Such measurements would greatly improve the quantitative constraints. Finally, we look forward to the results of NASA's upcoming Discovery-class mission to Psyche, which will yield new insight into the composition and origins of this unusual object.
\clearpage
\section*{Acknowledgements}
The authors are grateful to the NRAO and to Melissa Hoffman, Erica Keller, and Tony Remijan at the North American ALMA Science Center (NAASC) for their support in calibrating and imaging the ALMA data through a PI data reduction visit. KdK thanks Bryan Butler, Arielle Moullet, and Mike Brown. SC thanks Guy Consolmagno and Robert Macke at the Vatican Observatory for discussion about the thermophysical nature of Psyche's meteorite analogs, and Marco Delbo at the Observatoire de la C\^{o}te d'Azur for technical support with the thermophysical model. This paper makes use of the following ALMA data: ADS/JAO.ALMA\#2018.1.01271.S. ALMA is a partnership of ESO (representing its member states), NSF (USA) and NINS (Japan), together with NRC (Canada), MOST and ASIAA (Taiwan), and KASI (Republic of Korea), in cooperation with the Republic of Chile. The Joint ALMA Observatory is operated by ESO, AUI/NRAO and NAOJ. The National Radio Astronomy Observatory is a facility of the National Science Foundation operated under cooperative agreement by Associated Universities, Inc.
\begin{table}[h!]
\begin{center}
\caption{Observations and Derived Properties \label{tbl:obs}}
\begin{tabular}{ccccccccc}
Observation \# & T$_{start}^a$ & T$_{end}$ & t$_{int}$ & Sub-Obs Lon$^b$ & F$_{tot}^c$ & $<T_b>$ & $T_{b,peak}$ & Beam Size \\
 & [UT] & [UT] & [sec] & [deg] & [mJy] & [K] & [K] & [mas $\times$ mas] \\
\hline
1 & 06:33:12 & 06:37:46 & 200 & 200 & 54.8 $\pm$ 0.3 & 90.3 $\pm$ 0.1 & 117.9 $\pm$ 0.9 & (20.0,20.5) \\ 
2 & 06:38:11 & 06:42:47 & 218 & 193 & 54.5 $\pm$ 0.2 & 90.4 $\pm$ 0.1 & 120.5 $\pm$ 0.8 & (20.0,20.5) \\ 
3 & 06:43:41 & 06:48:52 & 200 & 185 & 55.4 $\pm$ 0.3 & 89.0 $\pm$ 0.1 & 117.4 $\pm$ 0.9 & (20.0,20.2) \\ 
4 & 06:49:17 & 06:54:41 & 218 & 176 & 55.4 $\pm$ 0.2 & 88.1 $\pm$ 0.1 & 121.9 $\pm$ 0.9 & (20.0,20.2) \\ 
5 & 06:54:43 & 07:00:35 & 218 & 168 & 56.0 $\pm$ 0.2 & 91.1 $\pm$ 0.1 & 116.6 $\pm$ 0.9 & (19.8,20.2) \\ 
6 & 07:00:59 & 07:06:23 & 218 & 160 & 56.1 $\pm$ 0.3 & 92.1 $\pm$ 0.1 & 113.9 $\pm$ 0.9 & (19.7,20.4) \\ 
7 & 07:17:05 & 07:21:38 & 200 & 137 & 59.2 $\pm$ 0.3 & 94.2 $\pm$ 0.1 & 117.7 $\pm$ 1.0 & (19.4,20.4) \\ 
8 & 07:22:03 & 07:25:36 & 163 & 131 & 60.1 $\pm$ 0.3 & 96.3 $\pm$ 0.1 & 119.0 $\pm$ 1.0 & (19.4,20.5) \\ 
9 & 07:25:45 & 07:31:26 & 200 & 124 & 60.4 $\pm$ 0.3 & 93.9 $\pm$ 0.1 & 119.3 $\pm$ 0.9 & (19.3,20.4) \\ 
10 & 07:31:50 & 07:35:23 & 164 & 117 & 61.9 $\pm$ 0.3 & 97.7 $\pm$ 0.1 & 123.5 $\pm$ 1.0 & (19.1,20.4) \\ 
11 & 07:35:48 & 07:40:13 & 181 & 111 & 62.3 $\pm$ 0.3 & 95.2 $\pm$ 0.1 & 121.9 $\pm$ 1.1 & (18.9,20.4) \\ 
12 & 08:02:30 & 08:07:43 & 200 & 72 & 64.8 $\pm$ 0.3 & 91.4 $\pm$ 0.1 & 128.9 $\pm$ 0.9 & (18.9,22.1) \\ 
13 & 08:08:08 & 08:13:35 & 218 & 64 & 64.5 $\pm$ 0.3 & 93.0 $\pm$ 0.1 & 121.0 $\pm$ 0.9 & (19.0,22.0) \\ 
14 & 08:13:37 & 08:19:32 & 218 & 55 & 62.6 $\pm$ 0.3 & 94.2 $\pm$ 0.1 & 118.5 $\pm$ 0.9 & (19.1,21.9) \\ 
15 & 08:19:58 & 08:25:28 & 218 & 47 & 60.8 $\pm$ 0.3 & 94.8 $\pm$ 0.1 & 119.5 $\pm$ 0.9 & (19.0,21.8) \\ 
16 & 08:25:55 & 08:31:05 & 200 & 38 & 58.1 $\pm$ 0.3 & 95.1 $\pm$ 0.1 & 115.9 $\pm$ 0.9 & (19.3,22.1) \\ 
17 & 08:31:31 & 08:36:22 & 182 & 31 & 56.7 $\pm$ 0.3 & 94.2 $\pm$ 0.1 & 115.4 $\pm$ 0.9 & (19.2,21.9) \\ 
18 & 08:47:05 & 08:51:42 & 200 & 9 & 54.0 $\pm$ 0.3 & 88.5 $\pm$ 0.1 & 116.7 $\pm$ 0.9 & (19.3,21.9) \\ 
19 & 08:52:08 & 08:56:28 & 200 & 2 & 54.1 $\pm$ 0.3 & 88.0 $\pm$ 0.1 & 121.6 $\pm$ 0.9 & (19.4,21.8) \\ 
20 & 08:57:24 & 09:02:37 & 200 & 353 & 54.4 $\pm$ 0.2 & 89.8 $\pm$ 0.1 & 116.9 $\pm$ 0.8 & (19.5,21.8) \\ 
21 & 09:03:03 & 09:06:38 & 163 & 346 & 55.7 $\pm$ 0.3 & 89.7 $\pm$ 0.1 & 116.4 $\pm$ 1.0 & (19.6,21.8) \\ 
22 & 09:07:36 & 09:11:13 & 163 & 340 & 54.8 $\pm$ 0.3 & 88.0 $\pm$ 0.1 & 115.6 $\pm$ 1.0 & (19.7,21.8) \\ 
\end{tabular}
\end{center}
$^a$Times given in UT on June 19, 2019.\\
$^b$Sub-observer longitude of Psyche at the midpoint time of observation from the \cite{shepard2021} model, corrected for light travel time. The sub-solar longitudes are 11$^{\circ}$ larger than the sub-observer longitudes; sub-observer and sub-solar latitudes are -14$^{\circ}$ and 3$^{\circ}$ (respectively) for all observations; and the solar phase angle is 17$^{\circ}$. \\
$^c$Total flux density of Psyche. The 3$\sigma$ uncertainty is presented; it is appropriate for comparison between observations, but does not include the flux density scale calibration uncertainty of 3\% that applies to the absolute calibration of the dataset as a whole.
\end{table}
\noindent
\begin{table}
    \caption{Candidate scenarios for Psyche's surface composition\label{tbl:scenarios}\\}
    \hskip-1.5cm
    \begin{tabular}{p{3cm}|p{5cm}|p{7cm}|p{3cm}}
    \hline
    \textbf{Surface Composition} & \textbf{Relevant Properties} & \textbf{Consistent? Under what conditions?} & \textbf{Implied surface density$^a$}\\
    \hline
    \hline
    
Pure silicate & 100\% silicate & No: The observed emissivity is too low regardless of specific composition, texture, or porosity: Figure \ref{fig:dielectricmodel_LLL_CI}. & N/A \\
\hline
Pure metal oxide/sulfide & 100\% metal in oxides/sulfides & Possible: The observed emissivity can be achieved with 40\% porosity (Figure \ref{fig:dielectricmodel_LLL_CI}a) but the polarization favors a lower-loss matrix. & 2.8 g/cm$^3$ \\
\hline
Pure metal & 100\% metal in metallic form & Unlikely: The observed emissivity is too high for the surface to be solid, conducting metal. Relevant lab data do not exist for powdered metals. & N/A \\
\hline
Pallasite/ mesosiderite & Surface is a mixture of metals and olivine/pyroxene & Possible: The observed emissivity can be achieved with a porosity in the 20-75\% range and metal content of 25-45\%, but the polarization favors a lower-loss matrix than the Fe-rich silicates in these meteorite classes. & 1.3-3.5 g/cm$^3$ depending on metal content \\
\hline
Enstatite chondrite & Surface is 65\% low-iron silicate and 35\% metal, of which 1/3 is FeS and 2/3 is metallic (using the example of Indarch) & Yes: for a porosity of 28\%: Figure \ref{fig:dielectricmodel_LLL_CI}c & 3.2 g/cm$^3$\\
\hline
CB chondrite & Surface is 50\% metal in the form of Fe-Ni and 50\% enstatite/olivine & Yes: for a porosity of 75\%: Figure \ref{fig:dielectricmodel_LLL_CI}b. The polarization favors low-iron silicates. & 2.0 g/cm$^3$\\
\hline
Silicate surface with solidified ferrovolcanic flows & Silicate and metal regions spatially interspersed & Possible: Emissivity is consistent if regions are interspersed below the $\sim$10 km scale and pure metal constitutes $\sim$40\% of the surface area, but polarization is difficult to explain. & Wide range possible \\ 
\hline
\end{tabular}
    
\footnotesize{$^a$Computed assuming a density of 4.6 g/cm$^3$ for iron sulfides, 7.8 g/cm$^3$ for metal, and 3.2 g/cm$^3$ for rock. The value is computed for the porosity given in the third column.}
\end{table}
\clearpage
\bibliography{psyche}{}
\bibliographystyle{aasjournal}

\end{document}